\RequirePackage{fix-cm}

\documentclass[smallcondensed]{svjour3}

\smartqed

\usepackage[utf8]{inputenc}
\usepackage{tabularx}
\usepackage{booktabs}
\usepackage{soul}
\usepackage{xcolor}
\usepackage{xspace}
\usepackage{graphicx}
\usepackage[caption=false]{subfig}  
\usepackage[export]{adjustbox}
\usepackage{amsmath}
\usepackage{textcomp}
\usepackage{listings}
\usepackage{algorithm}
\usepackage[noend]{algpseudocode}
\usepackage{balance}
\usepackage{enumitem}
\usepackage{multirow}
\usepackage{colortbl}
\usepackage{threeparttable}
\usepackage{multicol}
\usepackage{tcolorbox}
\usepackage{array}
\usepackage{longtable}
\usepackage{amssymb}
\usepackage[authoryear]{natbib}
\usepackage[english]{babel}
\usepackage[autostyle, english = american]{csquotes}
\usepackage[T1]{fontenc}
\usepackage{lmodern} 
\usepackage{microtype}
\usepackage{textgreek}
\usepackage{caption}
\usepackage[colorlinks=true, allcolors=black]{hyperref}
\usepackage[shortcuts,acronym]{glossaries}
\newacronym{hax}{HAX}{Human-AI Experience}
\newacronym{ide}{IDE}{Integrated Development Environment}
\newacronym{hci}{HCI}{Human-Computer Interaction}
\newacronym{sdlc}{SDLC}{Software Development Lifecycle}
\newacronym{in-ide-hax}{in-IDE HAX}{Human-AI Experience in Integrated Development Environment}
\newacronym{ai}{AI}{Artificial Intelligence}
\newacronym{llm}{LLM}{Large Language Model}
\definecolor{gray}{gray}{0.9}
\newcommand{\ie}{\emph{i.e.,}\xspace}
\newcommand{\eg}{\emph{e.g.,}\xspace}
\newcommand{\change}[1]{{\color{black}#1}}
\glsdisablehyper

\begin{document}

\title{Human-AI Experience in Integrated Development Environments: A Systematic Literature Review}

\titlerunning{in-IDE HAX: Literature Review}

\author{Agnia Sergeyuk         \and
        Ilya Zakharov \and
        Ekaterina Koshchenko \and
        Maliheh Izadi \and
}

\institute{A. Sergeyuk \at
              JetBrains Research, Belgrade, Serbia \\
              \email{agnia.sergeyuk@jetbrains.com}           
           \and
           I.Zakharov\at
              JetBrains Research, Belgrade, Serbia \\
              \email{ilia.zakharov@jetbrains.com}           
           \and
           E. Koshchenko\at
              JetBrains Research, Amsterdam, The Netherlands \\
              \email{ekaterina.koshchenko@jetbrains.com}           
           \and
           M. Izadi\at
              Delft University of Technology, Delft, The Netherlands \\
              \email{m.izadi@tudelft.nl}       
}

\date{Received: date / Accepted: date}

\maketitle

\begin{abstract}
The integration of \ac{ai} into \acp{ide} is reshaping software development, fundamentally altering how developers interact with their tools. This shift marks the emergence of \ac{in-ide-hax}, a field that explores the evolving dynamics of Human-Computer Interaction in \ac{ai}-assisted coding environments. Despite rapid adoption, research on \ac{in-ide-hax} remains fragmented, which highlights the need for a unified overview of current practices, challenges, and opportunities. To provide a structured overview of existing research, we conduct a systematic literature review of 90 studies, summarizing current findings and outlining areas for further investigation. 

We organize key insights from reviewed studies into three aspects: Impact, Design, and Quality of \ac{ai}-based systems inside \acp{ide}. \change{Impact findings show that \ac{ai}-assisted coding enhances developer productivity but also introduces challenges, such as verification overhead and over-reliance. Design studies show that effective interfaces surface context, provide explanations and transparency of suggestion, and support user control. Quality studies document risks in correctness, maintainability, and security. For future research, priorities include productivity studies, design of assistance, and audit of \ac{ai}-generated code. The agenda calls for larger and longer evaluations, stronger audit and verification assets, broader coverage across the software life cycle, and adaptive assistance under user control.}

\end{abstract}

\section{INTRODUCTION}

The integration of Artificial Intelligence (\ac{ai}) into Integrated Development Environments (\acp{ide}) has gained significant traction in recent years, driven by the promise of enhanced developer productivity, improved software quality, and more efficient workflows~\citep{ziegler2022productivity,izadi2024language}.
Recent large-scale industry surveys show that 76\% of respondents are either using or planning to use \ac{ai} tools in their development process~\citep{StackOverflow2024AI}.
Additionally, more than 50\% of respondents note such benefits as reduced time spent searching for information, faster coding and development, quicker completion of repetitive tasks, and an overall increase in productivity~\citep{JetBrains2024AI}.

Traditionally, \ac{hci} has focused on how users interact with software and tools. 
With \ac{ai} now integrated into development workflows, this focus is shifting toward \ac{hax}.
Unlike conventional software, \ac{ai} acts as an active collaborator rather than just a tool, changing how developers work and interact with their environment.

Given the growing reliance on \ac{ai}-enhanced \acp{ide} in software engineering (\eg Coursor\footnote{Cursor: The \ac{ai} Code Editor \url{https://www.cursor.com/}}), it is crucial to merge and evaluate existing evidence in the field of \ac{in-ide-hax}.
Doing so will help to identify recurring themes and gaps in the literature regarding the developer experience with \ac{ai} assistance.

\change{Recent literature reviews have provided comprehensive overviews of how \ac{llm} support software engineering tasks: summarizing model capabilities, task coverage, and evaluation techniques~\citep{he2025llm,zhou2025large,hou2024large,durrani2024decade}. However, these reviews do not analyze the interaction experience inside developer workspaces, where the software development lifecycle takes place. Our work is, to the best of our knowledge, the first systematic review of \ac{in-ide-hax}. This lens highlights experience-level patterns and design challenges that general and capability-focused reviews on \ac{llm} in software engineering do not capture.

This work builds on our earlier survey of \ac{in-ide-hax}~\citep{sergeyuk2024ide}, which provided an initial exploration of the topic and highlighted its potential. In contrast, the present study is extended in both scope and methodological rigor. It applies a formal Systematic Literature Review (SLR) process, following the Preferred Reporting Items for Systematic reviews and Meta-Analyses (PRISMA) framework~\citep{moher2010preferred}. Compared to the 36 papers in the initial study, which did not follow a systematic protocol, it includes 90 papers identified through a multi-step procedure that combines database search, backward snowballing, and expert-informed additions. Beyond expanding the dataset, this review introduces several analytical contributions. It not only organizes findings along three key research dimensions: Design, Impact, and Quality. It also provides a detailed analysis of the study context, empirical methods, and proposed future directions. These contributions position this study not as a follow-up review, but as a foundational reference for cumulative and reproducible progress in the research field of \ac{in-ide-hax}.

Using our protocol, we identified 257 studies and thoroughly reviewed 90 that were relevant to \ac{in-ide-hax}.
The goals of this review are to (a) identify common themes and patterns in the literature, (b) merge the current state of knowledge regarding \ac{in-ide-hax}, and (c) identify critical gaps that can guide future research efforts. We focus on how the presence of an \ac{ai} assistance changes what happens in the very place where code is written, inspected, or tested, rather than how AI helps in software engineering broadly.

Our review reveals the focus of \ac{in-ide-hax} research activity on professional software development settings, examining the code implementation stage of the \ac{sdlc}, often using GitHub Copilot as the primary example. Other lifecycle stages, such as requirements, testing, and deployment, receive limited attention. Educational contexts and perspectives of non-users or people who have stopped using \ac{ai} assistance are also rarely studied.

According to our review, key findings in the field span three core dimensions: Impact (effects on developers, tasks, and workflows), Design (how AI is integrated into the IDE and interacted with), and Quality (properties of AI outputs such as correctness, security, and readability). \emph{Impact}-oriented studies (74/90) report productivity improvements, especially among experienced users, but also note increased time spent on result verification and concerns about over-reliance. Studies in the \emph{Design} dimension (28 out of 90 papers) examine autocompletion and conversational systems, with some exploring emerging hybrid approaches. Research highlights the influence of prompt structure and interface properties such as context awareness, transparency, and user control on experience with \ac{ai} tooling. In the \emph{Quality} dimension (19/90), studies evaluate correctness, maintainability, and security of generated code, often calling for improved validation and diagnostic support.

Future research directions identified in the literature focus on several key areas, including productivity factors, audit mechanisms for AI-generated code, and the design of AI assistance tools. There is a strong emphasis on personalization and transparency, with a growing interest in exploring the long-term adoption of AI tools. Methodologically, studies suggest the need for larger-scale evaluations, comparative research across user groups and contexts, and longitudinal studies to track the evolution of AI's impact over time. Despite significant progress, many research directions remain underexplored, particularly in areas such as AI governance, user control, and proactivity. 

In summary, while substantial progress has been made in understanding in-IDE HAX, key gaps remain, particularly in exploring non-user perspectives and underrepresented stages of the software development lifecycle. Future research should focus on broadening the contexts studied, incorporating longitudinal designs, and refining methodologies to provide a more comprehensive understanding of AI's long-term impact and its role across diverse user groups.}

Our study offers the following contributions:

\begin{itemize}
    \item \textbf{Comprehensive synthesis of existing research.} We analyze 90 studies, categorizing them based on their research goals, methodological approaches, and key findings. This synthesis reveals dominant research themes, study contexts, and trends in \ac{hax} research.

    \item \textbf{Characterization of \ac{hax} research dimensions.} We classify research efforts into three core aspects: (a) \textbf{Impact}, examining how \ac{ai} affects developers' workflows, productivity, and experience; (b) \textbf{Design}, focusing on how \ac{ai} assistance is integrated into \acp{ide};  and (c) \textbf{Quality}, assessing \ac{ai}-generated code for security, comprehensibility, and adequacy.

    \item \textbf{Identification of future research opportunities.} By analyzing the proposed future work in the studies, we identify under-explored topics, such as \change{AI governance, user control, and proactivity}.

\end{itemize}

By addressing these aspects, we provide a structured overview of \ac{in-ide-hax} research, informing both academic and industry efforts to design and evaluate \ac{ai}-powered development tools.

\section{METHOD} \label{sec:Method}
To answer our research questions, we set out to conduct a systematic literature review in accordance with the Preferred Reporting Items for Systematic Reviews and Meta-Analyses (PRISMA) framework~\citep{moher2010preferred}.

The following Research Questions (RQs) guided our literature review:
\begin{itemize}  
    \item \textbf{RQ 1. What aspects of software development related to HAX within IDEs have been extensively studied, and which areas remain under-explored?}	
    To address RQ1, we defined the stages of \ac{sdlc} according to the Waterfall model~\citep{alshamrani2015comparison}.
    The Waterfall model was chosen for its clear delineation of development stages, which facilitated systematic categorization. 
    We also examined study contexts and three key aspects of \ac{hax} identified by a previous literature review~\citep{sergeyuk2024ide}.

    \item \textbf{RQ 2. What are the key findings in the field of \ac{hax} in \acp{ide}?}
    To answer RQ2, we systematically examined goals, research questions, and key findings from published reports.
    That allowed us to determine key research themes in the field and define methodological trends.
    
    \item \textbf{RQ 3. What are the potential directions for future research and development in the field of \ac{hax} in \acp{ide}?}
    To address RQ3, we analyzed the lists of future research directions proposed in the reviewed studies, identifying recurring themes and uncovering how studies build on each other's findings.

\end{itemize}

\change{To address RQ1 and RQ2, we applied an in-IDE HAX taxonomy~\citep{sergeyuk2024ide} developed through a staged qualitative procedure. Two authors independently conducted open coding of the corpus to propose candidate categories, followed by consensus meetings to merge or split codes, assign names, and draft a codebook with definitions and examples. The codebook was piloted on a subset, refined, and then used to code the full dataset by two authors, with disagreements adjudicated by a third. Categories with low support or conceptual overlap were consolidated to improve parsimony while preserving coverage. The resulting structure comprises three distinct dimensions: \emph{Impact} (effects on developers, tasks, and workflows), \emph{Design} (how AI is integrated into and interacted with in the IDE), and \emph{Quality} (properties of AI outputs such as correctness, security, and readability).}

\subsection{Planning} \label{subsec: Planning}
Based on the best practices established in the field~\citep{moher2010preferred, zhou2015quality, carrera2022conduct, kitchenham2010systematic, kitchenham2009systematic}, we defined eligibility criteria to ensure a comprehensive and methodologically sound review. These criteria were guided by the research objectives and scope of this review. 
We report our protocol details below:
\begin{itemize}
\label{list:report chars}
    \item \textbf{Timeframe}: Studies published between 2022 and 2024 were included to ensure relevance to the advancements in \ac{ai}-driven IDE features.
    \item \textbf{Language}: Only studies published in English were included.
    \item \textbf{Publication Status}: The review included peer-reviewed journal articles and conference proceedings. Recognizing the need to account for the rapid development in this field, preprints and unpublished dissertations were also considered if they provided unique insights directly relevant to the research objectives.
    \item \textbf{Population}: We considered studies in the field of computer science that focus on software developers or users engaging with \ac{ai}-enhanced tools within \acp{ide}.
    \item \textbf{Intervention}: Studies were included if they investigated \ac{in-ide-hax}.
    \item \textbf{Comparison}: Not applicable.
    \item \textbf{Outcomes}: This review included studies that presented empirical findings, whether qualitative, quantitative, or mixed-methods.
    \item \textbf{Study Design}: The review included original empirical studies.
\end{itemize}

\change{\ac{ide} in this review denotes any workspace that simultaneously offers (a) code editing, (b) immediate execution or preview, and (c) contextual AI feedback in the same window. Traditional desktop IDEs (IntelliJ, VS Code) and notebook-style environments (Jupyter, Databricks) satisfy all three criteria; plug-ins that embed a full editor and run code inline also qualify. Tools that push AI suggestions outside the primary editing surface (\eg web explainers, offline static analysers) are excluded.}

During the planning stage, as recommended by the guidelines~\citep{DARE1995}, we established quality assessment criteria for studies that would be included after the initial screening phase. Our quality assessment questions were inspired by the recommendations of~\citet{zhou2015quality}:
\begin{itemize} 
    \item \textbf{Reporting}: Is there a clear statement of the aims of the research?
    \item \textbf{Rigor}: Is the study design clearly defined?
    \item \textbf{Credibility}: Is there a clear statement of findings related to the aims of research?
    \item \textbf{Relevance}: Is the study of value for research or practice?
\end{itemize}

Each question was accompanied by the following scale: 0 — No, and not considered; 0.5 — Partially; 1 — Yes.
Following recommended guidelines, we set the cutoff for further inclusion in the review at a score higher than 2.
These criteria ensured a systematic evaluation of the research's aims, study design, methodological soundness, value and validity of findings.

Furthermore, the data extraction form was defined during the planning stage.
This form was guided by our research questions and the goal of assessing the representativeness of the included studies, given their varying venues and scopes.
Therefore, in addition to the name of each report, the extraction form included the following fields;
authors names and affiliation, DOI, publication date, venue, goal of the study, research questions, key findings, future work, scope (based on the taxonomy defined in previous work~\citep{sergeyuk2024ide}), and \ac{sdlc} stage.

To ensure coverage of a diverse range of scholarly sources, we selected several well-known digital libraries for our initial search. 
These included \textit{ACM Digital Library, DBLP, IEEE Digital Library, ISI Web of Science, ScienceDirect, Scopus, Springer Link, and arXiv}. 
We included \textit{arXiv} to capture insights from emerging fields, recognizing that valuable information can often be found in non-peer-reviewed papers. 
Additionally, given the prevalence of positive results in published literature, we included preprints and unpublished dissertations to reduce the risk of excluding studies with negative or null findings.

\subsection{Identifying and Screening}

\change{In November 2024, after finishing the planning phase, we searched the aforementioned sources, using titles, keywords, and abstracts as the basis for our queries. We optimized for high recall at the search stage and enforced precision during manual screening. 

The search string has a two-part structure: a core query covering terminology related to AI assistants and Human-AI Experience (\texttt{S\textsubscript{core}}), optionally combined with a query specifying IDE-related terms (\texttt{S\textsubscript{ide}}).

\begin{itemize}
  \item \textbf{S\textsubscript{core}}: \textit{``AI Assistant'' OR ``AI Companion'' OR ``AI-Powered Programming Tool'' OR ``Code Completion Tool'' OR ``Coding Assistant'' OR ``Copilot'' OR ``Intelligent Code Assistant'' OR ``LLM-Based Coding Assistant'' OR ``LLM-Powered Coding Assistant'' OR ``LLM4Code'' OR ``Programming Assistant'' OR ``Human-AI Experience'' OR ``Human-AI Co-Creation'' OR ``Human-AI Collaboration'' OR ``Human-AI Interaction''}
  \item \textbf{S\textsubscript{ide}}: \textit{``Integrated Development Environment'' OR ``Code Editor'' OR ``Coding Environment'' OR ``Development Environment'' OR ``Programming Environment''}
\end{itemize}

The full query was used in IEEE Xplore, Web of Science, ScienceDirect, Scopus, Springer Link, and arXiv. In ACM Digital Library and DBLP, we used only \texttt{S\textsubscript{core}} due to limited filtering capabilities and higher baseline relevance. We did not include “developer” as a required keyword, since it tended to retrieve general developer studies and exclude relevant work focused on tools or interfaces that met our criteria but were not indexed using that term. All search string variations and their application per library are available in the Supplementary materials~\citep{hax_dataset_2025}. 

We complemented our database search with two additional inclusion strategies. First, we included 35 studies from our prior literature survey~\citep{sergeyuk2024ide}, which, although not systematic, provided coverage of early work in this fast-moving field. Second, we manually added 30 relevant studies known to the authors through domain expertise. These were identified as frequently cited or influential in recent research but were missed due to inconsistent metadata, terminology drift, or indexing limitations in digital libraries. This expert-based supplementation is recommended in SLR methodology~\citep{wohlin2014guidelines, kitchenham2009systematic} to reduce omission of contextually critical work. All manually included studies were subjected to the same quality assessment and eligibility criteria as search-derived papers to ensure methodological consistency.

All retrieved records were manually screened against our in-IDE HAX inclusion criterion. This step preserved recall while restoring precision.} In total, 223 papers entered the screening phase of the systematic literature review.
We identified and excluded 30 duplicates appearing across multiple databases.
Next, we conducted an initial screening, applying the exclusion criteria and identifying 114 papers out of the scope of the current SLR.

\subsection{Assessing Eligibility} \label{subsec: Eligibility}
In the next stage of the systematic literature review, we assessed the eligibility of 79 full-text articles based on their quality.
As described in Subsection~\ref{subsec: Planning}, we applied four evaluation criteria: Reporting, Rigor, Credibility, and Relevance.
Articles were accepted only if their combined score on these criteria exceeded 2. 

The quality assessment was conducted independently by the second and third authors, followed by a discussion to reconcile any discrepancies.
In cases of disagreement, the first author acted as an arbitrator.
This structured process ensured consensus among the reviewers for the final quality assessment.
A total of 8 papers were excluded due to insufficient quality scores \change{lower or equal to 2, dictated by insufficient relevance and rigor. Excluded papers were accessed by authors as only partially or not sufficient in terms of clarity of the statement of the aims of the research, its design, findings related to the aims of the research, and value for research or practice.}

\subsection{Backward Snowballing}
To complement the systematic search, we performed backward snowballing~\citep{wohlin2014guidelines} on papers deemed eligible for inclusion.
First, we extracted citations from these papers and screened them against the report-related criteria outlined in Subsection~\ref{subsec: Planning}.
References that did not meet these criteria were excluded.
We next removed any papers already identified in previous steps
We assessed the remaining studies against the study-level inclusion criteria described in Subsection~\ref{subsec: Planning}, excluding out-of-scope papers.
\change{After the first round of this process, 18 new in-scope papers were identified. A second round of snowballing did not yield any additional relevant papers.
These 18 studies underwent the eligibility assessment process outlined in Subsection~\ref{subsec: Eligibility}. 

During the revision phase, we re-verified the included studies. This process led to the removal of 2 papers (one out-of-scope, one duplicate) and the addition of 3 previously overlooked studies, ensuring that the final set reflects both the inclusion criteria and recent relevant literature and yielding a final total of 90 papers to be included in the review. 

Note that eligibility was determined by the date of first public availability. We included studies first available between January 2022 and November 2024, including preprints on arXiv. When a study later appeared in an archival venue in 2025, we cite the archival version with its DOI and venue year. As a result, some entries list 2025 in the reference list while remaining within the time window of our review based on the preprint date.}

\subsection{Data Extraction}
As mentioned in Subsection~\ref{subsec: Planning}, from all 90 eligible papers, we extracted data about \textit{Authors, Authors’ Affiliation, DOI, Publication Date, Venue, Goal of the Study, Research Questions, Key Findings, Future Work \change{with its topic and methodology}, Scope,  \ac{sdlc} Stage}.
In the later stages of data analysis, we added a \textit{Context} column to indicate whether \ac{hax} was studied in a professional or educational setting. \change{Moreover, we classified studies based on the types of tasks investigated in the work.}

This work was conducted manually by the first author, with support from an \ac{ai}-based system called Elicit~\footnote{Elicit: The \ac{ai} Research Assistant \url{https://elicit.com/}}.
This tool is designed to facilitate the extraction and summarization of scientific papers. It supported this work by formulating concise descriptions of each study's goal, key findings, and future work.
To ensure data integrity and accuracy, all tool-generated outputs were rigorously reviewed and cross-verified against the original papers.
Information was included into the final review only after confirming its faithful representation of the original meanings.
This hybrid approach ensured both efficiency and precision in the data extraction process.

Data extraction yielded a comprehensive table with 90 studies in it.
We provide the full table in Supplementary materials~\citep{hax_dataset_2025}, and present a shortened overview of the studies in Table~\ref{tab:overview}.

\small
\renewcommand{\arraystretch}{1.15}
\rowcolors{2}{gray!50}{white}
\setlength\LTcapwidth{\linewidth}
\begin{longtable}{@{}>{\raggedright\arraybackslash}p{0.02\textwidth}
                    >{\raggedright\arraybackslash}p{0.65\textwidth}
                    >{\raggedright\arraybackslash}p{0.25\textwidth}@{}}
\caption[Included studies]{List of studies included in the review.\\
\footnotesize\textit{Dating note.} \change{Inclusion used first public availability between January 2022 and November 2024. We cite the archival version with its DOI and venue year. Some entries show 2025 because an earlier preprint falls within this window.}\label{tab:overview}}\\

\toprule
\textbf{ID} & \textbf{Title} & \textbf{Authors} \\
\midrule
\endfirsthead
\toprule
\textbf{ID} & \textbf{Title} & \textbf{Authors} \\
\midrule
\endhead
\midrule
\multicolumn{3}{r}{\textit{Continued on next page}} \\
\midrule
\endfoot
\bottomrule
\endlastfoot
1 & An Empirical Evaluation of GitHub Copilot's Code Suggestions & \citet{nguyen2022empirical} \\
2 & \change{Asleep at the Keyboard? Assessing the Security of GitHub Copilot’s Code Contributions.} & \citet{pearce2022asleep} \\
3 & Assessing the quality of GitHub copilot’s code generation & \citet{yetistiren2022assessing} \\
4 & Better Together? An Evaluation of AI-Supported Code Translation & \citet{weisz2022better} \\
5 & Designing PairBuddy—A Conversational Agent for Pair Programming & \citet{robe2022designing} \\
6 & Discovering the Syntax and Strategies of Natural Language Programming with Generative Language Models & \citet{jiang2022discovering} \\
7 & Documentation Matters: Human-Centered AI System to Assist Data Science Code Documentation in Computational Notebooks & \citet{wang2022documentation} \\
8 & Expectation vs. Experience: Evaluating the Usability of Code Generation Tools Powered by Large Language Models & \citet{vaithilingam2022expectation} \\
9 & Exploring the Learnability of Program Synthesizers by Novice Programmers & \citet{jayagopal2022exploring} \\
10 & Github copilot in the classroom: learning to code with AI assistance & \citet{puryear2022github} \\
11 & How Readable is Model-generated Code? Examining Readability and Visual Inspection of GitHub Copilot & \citet{al2022readable} \\
12 & Investigating Explainability of Generative AI for Code through Scenario-based Design & \citet{sun2022investigating} \\
13 & Is GitHub Copilot a Substitute for Human Pair-Programming? An Empirical Study & \citet{imai2022github} \\
14 & Practitioners' expectations on automated code comment generation & \citet{hu2022practitioners} \\
15 & Productivity Assessment of Neural Code Completion & \citet{ziegler2022productivity} \\
16 & Taking Flight with Copilot: Early insights and opportunities of AI-powered pair-programming tools & \citet{bird2022taking} \\
17 & “It’s Weird That it Knows What I Want”: Usability and Interactions with Copilot for Novice Programmers & \citet{prather2023s} \\
18 & A Case Study in Engineering a Conversational Programming Assistant's Persona & \citet{ross2023case} \\
19 & A Case Study on Scaffolding Exploratory Data Analysis for AI Pair Programmers & \citet{zhou2023case} \\
20 & A Mixed Reality Approach for Innovative Pair Programming Education with a Conversational AI Virtual Avatar & \citet{manfredi2023mixed} \\
21 & AI-Powered Chatbots and the Transformation of Work: Findings from a Case Study in Software Development and Software Engineering & \citet{susse2023ai} \\
22 & Anticipating User Needs: Insights from Design Fiction on Conversational Agents for Computational Thinking & \citet{penney2023anticipating} \\
23 & Case Study: Using AI-Assisted Code Generation In Mobile Teams & \citet{vasiliniuc2023case} \\
24 & CoLadder: Supporting Programmers with Hierarchical Code Generation in Multi-Level Abstraction & \citet{yen2023coladder} \\
25 & Conversing with Copilot: Exploring Prompt Engineering for Solving CS1 Problems Using Natural Language & \citet{denny2023conversing} \\
26 & Copilot for Xcode: Exploring AI-Assisted Programming by Prompting Cloud-based Large Language Models & \citet{tan2023copilot} \\
27 & From ``Ban It Till We Understand It'' to ``Resistance is Futile'': How University Programming Instructors Plan to Adapt as More Students Use AI Code Generation and Explanation Tools such as ChatGPT and GitHub Copilot & \citet{lau2023ban} \\
28 & Grounded Copilot: How Programmers Interact with Code-Generating Models & \citet{barke2023grounded} \\
29 & How Novices Use LLM-Based Code Generators to Solve CS1 Coding Tasks in a Self-Paced Learning Environment & \citet{kazemitabaar2023novices} \\
30 & In-IDE Generation-based Information Support with a Large Language Model & \citet{nam2023ide} \\
31 & Is GitHub's Copilot as Bad as Humans at Introducing Vulnerabilities in Code? & \citet{asare2023github} \\
32 & Lost at C: A User Study on the Security Implications of Large Language Model Code Assistants & \citet{sandoval2023lost} \\
33 & On the Design of AI-powered Code Assistants for Notebooks & \citet{mcnutt2023design} \\
34 & On the Robustness of Code Generation Techniques: An Empirical Study on GitHub Copilot & \citet{mastropaolo2023robustness} \\
35 & Practices and Challenges of Using GitHub Copilot: An Empirical Study & \citet{zhang2023practices} \\
36 & Practitioners' Expectations on Code Completion & \citet{wang2023practitioners} \\
37 & Slide4N: Creating Presentation Slides from Computational Notebooks with Human-AI Collaboration & \citet{wang2023slide4n} \\
38 & Spellburst: A Node-based Interface for Exploratory Creative Coding with Natural Language Prompts & \citet{angert2023spellburst} \\
39 & \change{Studying the effect of AI Code Generators on Supporting Novice Learners in Introductory Programming} & \citet{kazemitabaar2023studying} \\
40 & The Impact of AI on Developer Productivity: Evidence from GitHub Copilot & \citet{peng2023impact} \\
41 & The Programmer’s Assistant: Conversational Interaction with a Large Language Model for Software Development & \citet{ross2023programmer} \\
42 & Towards More Effective AI-Assisted Programming: A Systematic Design Exploration to Improve Visual Studio IntelliCode’s User Experience & \citet{vaithilingam2023towards} \\
43 & Using GitHub Copilot to Solve Simple Programming Problems & \citet{wermelinger2023using} \\
44 & “It would work for me too”: How Online Communities Shape Software Developers’ Trust in AI-Powered Code Generation Tools & \citet{cheng2024would} \\
45 & A Large-Scale Survey on the Usability of AI Programming Assistants: Successes and Challenges & \citet{liang2024large} \\
46 & A Study on Developer Behaviors for Validating and Repairing LLM-Generated Code Using Eye Tracking and IDE Actions & \citet{tang2024study} \\
47 & A Transformer-Based Approach for Smart Invocation of Automatic Code Completion & \citet{de2024transformer} \\
48 & A User-centered Security Evaluation of Copilot & \citet{asare2024user} \\
49 & AI Tool Use and Adoption in Software Development by Individuals and Organizations: A Grounded Theory Study & \citet{li2024ai} \\
50 & An Analysis of the Costs and Benefits of Autocomplete in IDEs & \citet{jiang2024analysis} \\
51 & An Empirical Study of Code Search in Intelligent Coding Assistant: Perceptions, Expectations, and Directions & \citet{liu2024empirical} \\
52 & An Empirical Study on Usage and Perceptions of LLMs in a Software Engineering Project & \citet{rasnayaka2024empirical} \\
53 & An Exploratory Study on Upper-Level Computing Students' Use of Large Language Models as Tools in a Semester-Long Project & \citet{tanay2024exploratory} \\
54 & Analyzing Prompt Influence on Automated Method Generation: An Empirical Study with Copilot & \citet{fagadau2024analyzing} \\
55 & Ansible Lightspeed: A Code Generation Service for IT Automation & \citet{sahoo2024ansible} \\
56 & Are Prompt Engineering and TODO Comments Friends or Foes? An Evaluation on GitHub Copilot & \citet{obrien2024prompt} \\
57 & Can Developers Prompt? A Controlled Experiment for Code Documentation Generation & \citet{kruse2024can} \\
58 & Copilot Evaluation Harness: Evaluating LLM-Guided Software Programming & \citet{agarwal2024copilot} \\
59 & CoPrompt: Supporting Prompt Sharing and Referring in Collaborative Natural Language Programming & \citet{feng2024coprompt} \\
60 & Defendroid: Real-time Android code vulnerability detection via blockchain federated neural network with XAI & \citet{senanayake2024defendroid} \\
61 & Design Principles for Collaborative Generative AI Systems in Software Development & \citet{chen2024design} \\
62 & Developers' Perspective on Today's and Tomorrow's Programming Tool Assistance: A Survey & \citet{kuang2024developers} \\
63 & Evaluating Human-AI Partnership for LLM-based Code Migration & \citet{omidvar2024evaluating} \\
64 & Exploring Interaction Patterns for Debugging: Enhancing Conversational Capabilities of AI-assistants & \citet{chopra2024exploring} \\
65 & How Do Data Analysts Respond to AI Assistance? A Wizard-of-Oz Study & \citet{gu2024data} \\
66 & IDA: Breaking Barriers in No-code UI Automation Through Large Language Models and Human-Centric Design & \citet{shlomov2024ida} \\
67 & Identifying the Factors That Influence Trust in AI Code Completion & \citet{brown2024identifying} \\
68 & Improving Steering and Verification in AI-Assisted Data Analysis with Interactive Task Decomposition & \citet{kazemitabaar2024improving} \\
69 & Investigating and Designing for Trust in AI-powered Code Generation Tools & \citet{wang2024investigating} \\
70 & Investigating Interaction Modes and User Agency in Human-LLM Collaboration for Domain-Specific Data Analysis & \citet{guo2024investigating} \\
71 & Ivie: Lightweight Anchored Explanations of Just-Generated Code & \citet{yan2024ivie} \\
72 & Methodology for Code Synthesis Evaluation of LLMs Presented by a Case Study of ChatGPT and Copilot & \citet{sagodi2024methodology} \\
73 & Multi-line AI-assisted Code Authoring & \citet{dunay2024multi} \\
74 & Non-Expert Programmers in the Generative AI Future & \citet{feldman2024non} \\
75 & Performance, Workload, Emotion, and Self-Efficacy of Novice Programmers Using AI Code Generation & \citet{gardella2024performance} \\
76 & Prompt Sapper: A LLM-Empowered Production Tool for Building AI Chains & \citet{cheng2024prompt} \\
77 & Reading Between the Lines: Modeling User Behavior and Costs in AI-Assisted Programming & \citet{mozannar2024reading} \\
78 & Significant Productivity Gains through Programming with Large Language Models & \citet{weber2024significant} \\
79 & The RealHumanEval: Evaluating Large Language Models' Abilities to Support Programmers & \citet{mozannar2024realhumaneval} \\
80 & The Widening Gap: The Benefits and Harms of Generative AI for Novice Programmers & \citet{prather2024widening} \\
81 & Toward Effective AI Support for Developers: A survey of desires and concerns & \citet{khemka2024toward} \\
82 & Towards Feature Engineering with Human and AI’s Knowledge: Understanding Data Science Practitioners’ Perceptions in Human\&AI-Assisted Feature Engineering Design & \citet{zhu2024towards} \\
83 & Transforming Software Development: Evaluating the Efficiency and Challenges of GitHub Copilot in Real-World Projects & \citet{pandey2024transforming} \\
84 & Trust in Generative AI among students: An Exploratory Study & \citet{amoozadeh2024trust} \\
85 & Using AI Assistants in Software Development: A Qualitative Study on Security Practices and Concerns & \citet{klemmer2024using} \\
86 & Validating AI-Generated Code with Live Programming & \citet{ferdowsi2024validating} \\
87 & When to Show a Suggestion? Integrating Human Feedback in AI-Assisted Programming & \citet{mozannar2024show} \\
88 & \change{Exploring the Design Space of Cognitive Engagement Techniques with AI-Generated Code for Enhanced Learning} & \citet{kazemitabaar2025exploring} \\
89 & Exploring the problems, their causes and solutions of AI pair programming: A study on GitHub and Stack Overflow & \citet{zhou2025exploring} \\
90 & Generation Probabilities Are Not Enough: Uncertainty Highlighting in AI Code Completions & \citet{vasconcelos2025generation} \\
\end{longtable}
\normalsize

\section{RESULTS}

\begin{figure*}[tb]
    \centering
    \includegraphics[width=\linewidth]{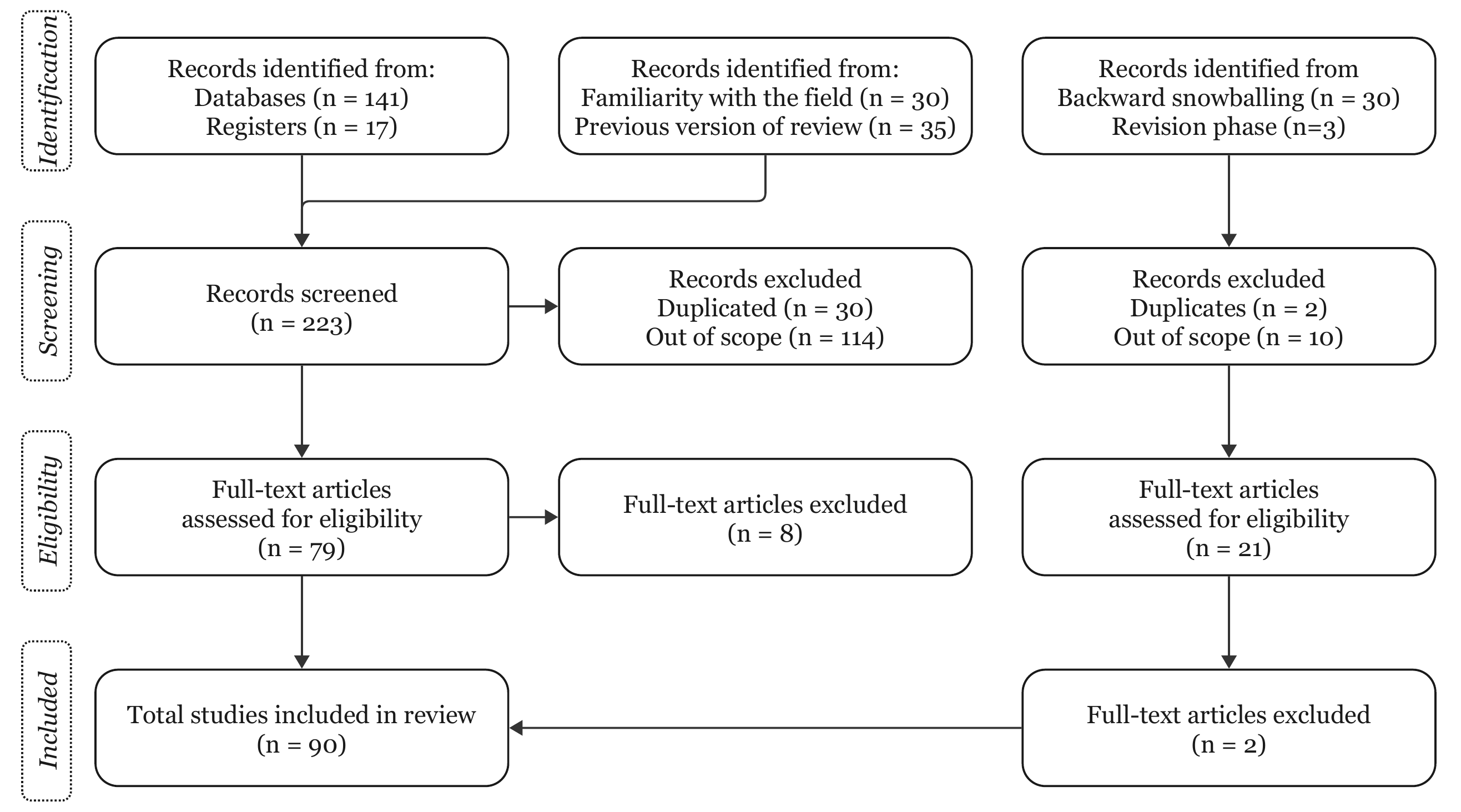}
    \caption{\change{Flow diagram of the study.}}
    \label{fig:scheme}
\end{figure*}

This section presents key findings from our review of the studies on \ac{in-ide-hax}. \change{We analyze the contexts of the studies, the reported impact of \ac{ai} on developers, the design of \ac{ai} tools for coding, and the quality of the \ac{ai} tools. The expanded corpus confirms the adequacy of the initial three-part framing~\citep{sergeyuk2024ide} and adds finer subscopes within each dimension, but it does not introduce any new top-level dimensions. We also examine future work directions proposed in the reviewed literature and methodological patterns of the studies. As described in Figure~\ref{fig:scheme} and Section~\ref{sec:Method}, we screened a total of 256 papers, 100 of which were assessed for eligibility, with 90 included in the final review stage.}

\subsection{Context of studies}

\begin{figure*}[tb]
    \centering
    \includegraphics[width=\linewidth]{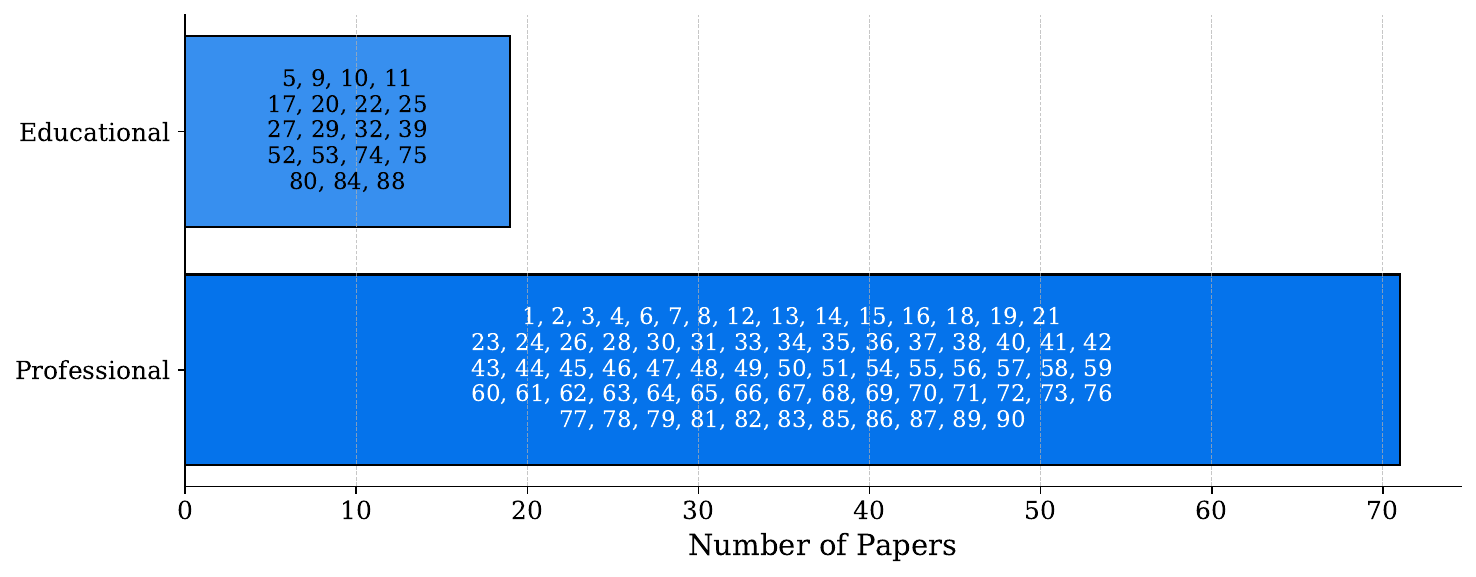}
    \caption{\change{Paper IDs by Context.}}
    \label{fig:context}
\end{figure*}

The studies analyzed in this review represent two primary contexts for working with \ac{ai} in software development: \textit{professional} \change{(71 out of 90) and \textit{educational} (19/90) (see Fig.~\ref{fig:context} for paper IDs)}. These contexts provide contrasting perspectives on the in-IDE integration of \ac{ai} tools, with professional studies focusing on real-world applications and educational studies exploring their role in learning environments. \change{Note that studies in educational settings may be underrepresented because our scope targets in-IDE contexts and students do not always work inside full IDEs during coursework. For a literature review of broader AI tools for coding in a strictly educational context, we refer the reader to ~\citep{agbo2025computing, ariza2025generative}}

The majority of professional studies emphasize the practical utility of \ac{ai} tools. Studies in this context address coding (\change{17/71}) and maintenance (\change{8/71}), with limited exploration of requirements gathering \change{(1/71). However, 46 out of 71 studies do not specify a target \ac{sdlc} stage, instead evaluating AI tooling for general programming use without stage-level framing, indicating a gap in contextual precision.} 

Educational studies primarily focus on pedagogical insights. These studies often investigate how students use \ac{ai} tools to learn programming or solve computational problems. \change{While a few studies address coding (6/16) and requirements formulation (1/16), most (13/16) do not specify the \ac{sdlc} stage, aligning with the broader exploratory nature of educational research.}

\begin{figure*}[tb]
    \centering
    \includegraphics[width=\linewidth]{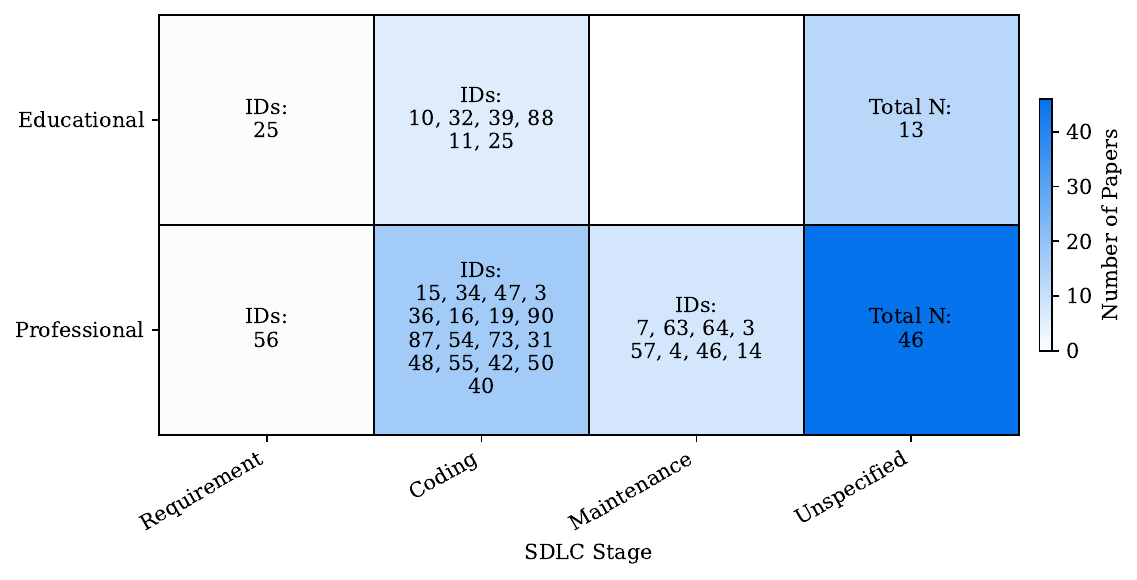}
    \caption{\change{Paper IDs by Context and SDLC stage.}}
    \label{fig:context_sdlc}
\end{figure*}

\change{The distribution depicted on Figure~\ref{fig:context_sdlc}, reveals notable gaps in the literature:
\begin{itemize}
    \item Lack of perspectives from people who do not use \ac{ai} tools for development for a variety of reasons.
    \item Limited number of educational studies highlighting the need for a deeper exploration of \ac{ai}’s impact on practical skills acquired during CS education, especially as developers constantly learn and re-learn programming practices in development environments.
    \item Lack of specification of the \ac{sdlc} stage, suggesting an opportunity for future research to contextualize \ac{ai} tool usage across all stages of the \ac{sdlc}.
\end{itemize}}

\change{\begin{figure*}[tb]
    \centering
    \includegraphics[width=\linewidth]{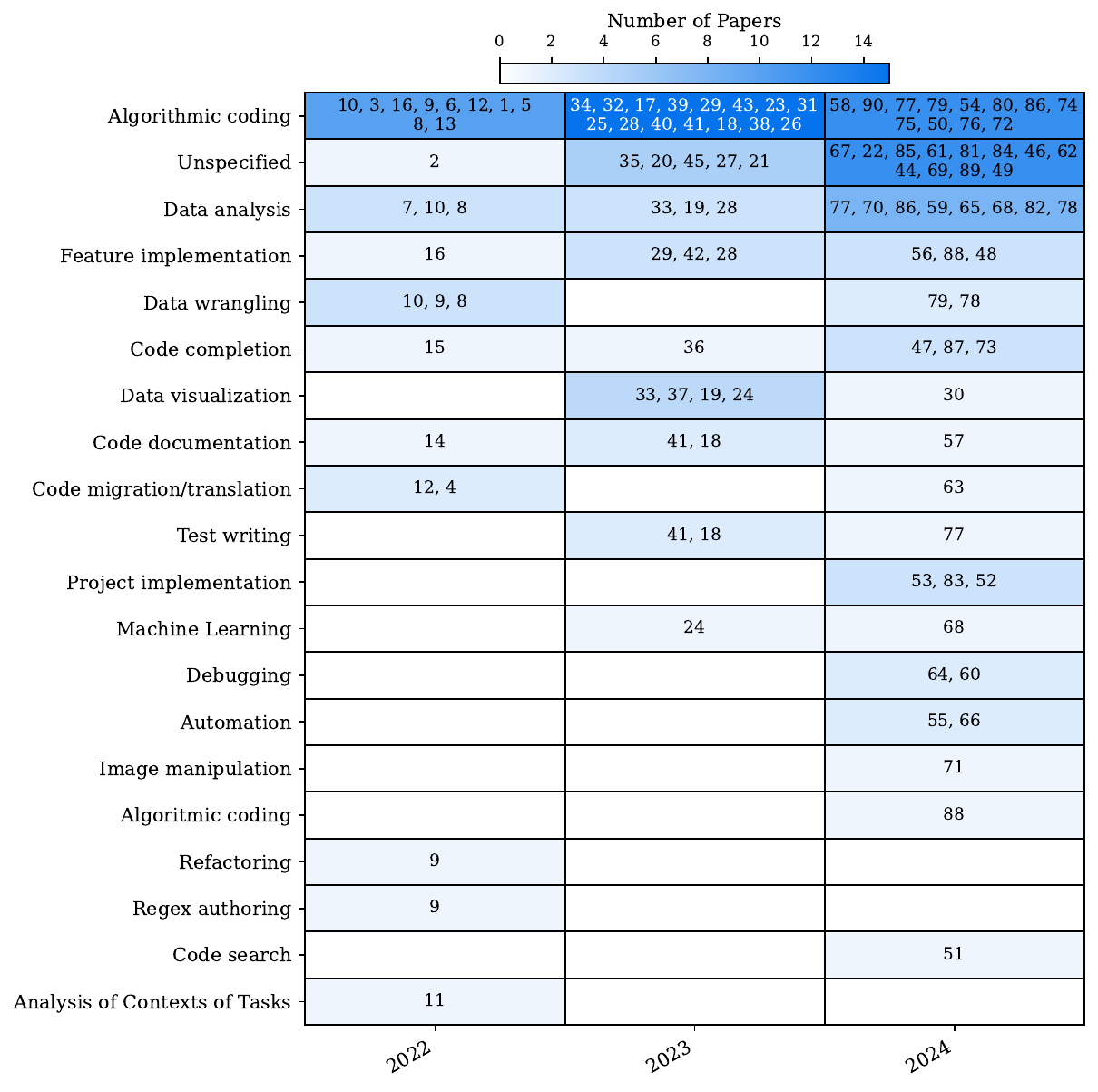}
    \caption{\change{Paper IDs by Types of tasks and Year.}}
    \label{fig:task_year}
\end{figure*}

Analysis of tasks assigned to the participants of studies reveals that they change over the three years. Early studies rely on smaller algorithmic exercises similar to LeetCode\footnote{LeetCode: a resource for technical interviews materials and coding competitions~\url{https://leetcode.com/}}
-type programming tasks. Later studies introduce tasks that require data analysis, or a project-level use of AI. Data analysis tasks become more common over time. Studies add wrangling, plotting, and notebook workflows, which reflects the maturity of AI support in these environments and the relevance of studying such use. This broadening is also visible in the growth of the \textit{Unspecified} category in Fig.~\ref{fig:task_year}, where authors evaluate AI assistance as general programming support without tying it to a single topic or SDLC stage. The variety of assigned tasks increases across years. The corpus now includes not only algorithmic coding and data analysis, but also documentation, testing, and a few cases of debugging and migration. This spread enables more precise claims about where AI helps, but also raises the need for clearer task descriptions to support comparison across studies.}

A subset of studies conceptualizes programming with \ac{ai} as a form of \textit{pair programming}, where \ac{ai} acts as a collaborative partner\change{~\citep{zhou2025exploring,imai2022github,zhou2023case,bird2022taking,kazemitabaar2023studying}}. This perspective highlights the evolving view of \ac{ai} as not just a tool but a collaborator, representing a growing trend in co-creative programming workflows and suggesting new opportunities and challenges for both practitioners and educators.

\begin{figure*}[tb]
    \centering
    \includegraphics[width=.9\linewidth]{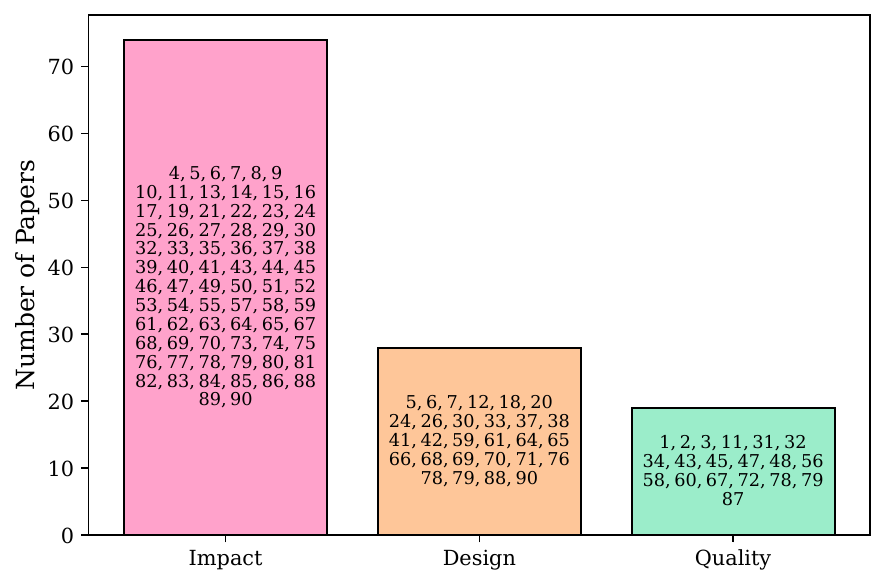}
    \caption{\change{Paper Counts and IDs by Scope.}}
    \label{fig:scope}
\end{figure*}

\subsection{Impact}

Most prominent branch of research is dedicated to \textit{Impact of \ac{hax} in \ac{ide}} with \change{74 out of 90} (see Fig.~\ref{fig:scope} for IDs) reviewed studies focusing on it. The increasing integration of \ac{ai}-powered tools in \acp{ide} has reshaped the way developers engage with coding, debugging, documentation, and software design\change{~\citep{peng2023impact,weber2024significant}}. \ac{ai}’s role extends beyond simple automation, influencing productivity and software engineering practices, \eg \change{~\citep{nguyen2022empirical, vaithilingam2022expectation}}. Although \ac{ai} has undoubtedly enhanced productivity in some contexts, its overall effects remain highly context-dependent and nuanced. Studies consistently highlight the dual nature of \ac{ai}'s impact, providing both benefits and challenges.

Multiple \change{studies looked into the productivity aspect of coding with AI (13/74)}. Some of them have confirmed a notable boost in productivity when developers incorporate \ac{ai} into their workflows. Potential benefits include a higher ``ceiling'' of productivity for experienced developers and faster onboarding for newcomers~\citep{vasiliniuc2023case}. For instance, participants using GitHub Copilot completed \change{implementation of an HTTP server in JavaScript}  up to 55.8\% faster than control groups~\citep{peng2023impact}, while other investigators observed productivity gains ranging from 26\% to 35\% \change{for complex tasks, across many files, and in proprietary contexts}~\citep{pandey2024transforming}. These improvements are partially attributed to fewer context switches and reduced boilerplate coding, enabling developers to offload repetitive tasks to \ac{ai} while focusing on higher-level logic. However, gains are inconsistent: \change{developers report that tasks involving proprietary or highly complex logic see only marginal benefits, as current assistance struggles with context-specific nuances of the real-world programming~\citep{khemka2024toward}.}

Although \ac{ai}-driven suggestions frequently accelerate development, they also require new verification efforts. When using \ac{ai} tools, verifying suggestions, refining prompts, and reworking code generated by \ac{ai} might take up to 50\% of developers’ time~\citep{mozannar2024reading}. This increased mental overhead arises because \ac{ai}’s output can be partially correct yet subtly flawed \citep{wermelinger2023using}, requiring careful review and, at times, stepwise interaction or re-prompting \citep{kazemitabaar2024improving}. \change{Recent work proposes mechanisms that reduce this cost, showing that highlighting tokens by predicted edit likelihood improves speed and directs attention to problematic regions~\citep{vasconcelos2025generation}.} Such extra validation often mitigates the risk of blindly accepting erroneous suggestions, \ie phenomenon known as ``automation bias''~\citep{al2022readable}.

\change{Studies focusing on novice programmers reveal both benefits and pronounced risks. Across classroom and lab settings, \ac{ai} assistance tends to speed novice work and reduce perceived workload while short-term learning remains intact when use is structured~\citep{kazemitabaar2023studying}. Beginners often solve \change{programming assignment} quicker with \ac{ai}'s help~\citep{puryear2022github,rasnayaka2024empirical,tanay2024exploratory}. Time-pressured studies report lower mental workload and steady self-efficacy with assistance, with additional time using \ac{ai} relating to better task outcomes~\citep{gardella2024performance}. However, beginners can over-trust \ac{ai}, losing sight of important foundational concepts~\citep{prather2024widening}. In certain cases, when system suggestions are misleading, novices drift away from correct solutions, as they lack the expertise to detect subtle errors~\citep{zhou2023case}. Structured prompts and explanatory interfaces can help novices understand \ac{ai} outputs better, fostering safer learning pathways~\citep{kruse2024can}. A hybrid strategy that mixes authoring and modification is associated with better results in self-paced CS1 tasks, and prompting is a learnable activity that raises solve rates when taught explicitly \citep{kazemitabaar2023novices,denny2023conversing}.}

\subsubsection{Attitude}

\change{A subset of studies (\change{11/74}) examines user attitude toward \ac{ai} assistance in \acp{ide}, with trust as the central construct. Trust is shaped by both the properties of the suggestions and the expertise of the user~\citep{amoozadeh2024trust,brown2024identifying}. High-quality and context-relevant suggestions raise trust, while inconsistency and unnecessary complexity lower it. Layers of outputs that help users judge suggestions, such as brief explanations, usage statistics, and control over scope, are proposed as practical supports for trust formation~\citep{wang2024investigating}. Context also matters. Trust drops for high-stakes or production-related work and for complex or open-ended tasks, and it rises for routine work or proof-of-concept code~\citep{wang2024investigating}. In addition, developers treat suggestions in test files more cautiously than those in production files, and they are less likely to accept code completion suggestions when writing tests~\citep{brown2024identifying}.

Attitude relates to reliance. Over-reliance appears when users accept \ac{ai} output without sufficient checking, and under-reliance appears when they dismiss correct output because justification feels costly. Novices often overestimate the reliability of \ac{ai} output and invest less time in verification for algorithmic problems~\citep{al2022readable}, so they tend to over-rely on AI. In educational settings, reliance and trust vary widely and depend on perceived performance of the system and task context, which warrants explicit trust calibration rather than blanket bans~\citep{amoozadeh2024trust,brown2024identifying,wang2024investigating,lau2023ban}. Students also renegotiate authorship and autonomy when assistants anticipate intent, which calls for assessment that values reasoning traces and verification~\citep{prather2023s}. Moreover, novices struggle to form accurate mental models of synthesizers, so interfaces should expose system state and allow incremental control~\citep{jayagopal2022exploring}. Professional developers sometimes under-rely and reject correct outputs that lack a clear rationale~\citep{sun2022investigating}. Tooling can mediate both patterns. Live programming environments that surface runtime values reduce the effort of review and support more calibrated acceptance decisions~\citep{ferdowsi2024validating}.}

\subsection{Design}

To account for the changes introduced by \ac{ai} integration in \acp{ide}, a focused body of research has developed on the \textit{Design} of \ac{hax} in \ac{ide}. Among the reviewed studies, \change{28 out of 90 (see Fig.~\ref{fig:scope} for IDs)} investigate forms of \ac{ai} integration into \acp{ide}, with most of them (22/28) explicitly examining how these forms of integration impact users, and 17 of them exploring design principles.

The research presents two primary paradigms of \ac{ai}-assisted interaction inside \acp{ide}: autocompletion-based assistance and conversational agents. Autocompletion interfaces provide short, low-friction code completions, enabling developers to work seamlessly with inline suggestions. These approaches minimize cognitive load and support rapid decision-making without disrupting workflow~\citep{vaithilingam2023towards, yen2023coladder}. In contrast, conversational interactions (often embodied in chat-based assistants or virtual avatars) facilitate higher-level reasoning, iterative refinement, and complex problem-solving~\citep{ross2023programmer, robe2022designing, gu2024data}. Rather than being opposing paradigms, these interaction modes are complementary~\citep{weber2024significant}, with studies advocating for hybrid models that seamlessly integrate autocompletion with interactive, context-sensitive conversations. In addition, some studies experiment with alternative modalities of interaction that go beyond these two dominant paradigms. For example, mixed-reality interfaces~\citep{manfredi2023mixed} introduce conversational \ac{ai} in augmented pair-programming environments, leveraging immersive learning experiences.

Due to the conversational nature of interaction with \ac{ai} in \ac{ide}, prompting became a prominent object of investigations with \change{7 out of 28} studies focusing on this aspect. Across reviewed studies, prompt engineering is positioned as a strategic skill that shapes how developers and \ac{ai} systems collaborate, affecting both individual productivity and team-based workflows\change{~\citep{kruse2024can,yen2023coladder}. There are claims that prompting should be taught as a skill, with scaffolds for decomposition, short rationales before revealing solutions, and reusable prompt solution pairs delivered in step-by-step dialogues ~\citep{kazemitabaar2025exploring}}. A central challenge in \ac{ai}-assisted development is ensuring that \ac{ai}-generated code aligns with user intent\change{~\citep{fagadau2024analyzing, denny2023conversing,obrien2024prompt}}. Therefore, studies investigating prompt formulation demonstrate that structured and well-contextualized prompts lead to more accurate and reliable output, while excessive detail can degrade performance. As \ac{ai} tools become embedded in professional and educational workflows, collaborative prompt engineering is also gaining importance\change{~\citep{feng2024coprompt, cheng2024prompt}}. Consequently, authors introduce frameworks for shared prompt refinement, knowledge transfer, and workflow optimization, suggesting the need for modularity and re-usability in \ac{ai} interactions.

\change{For education, in-IDE \ac{hax} should shift learners from code consumption to deliberate practice. Interfaces need small, purposeful friction before code adoption, which would improve transfer and calibration without added load~\citep{kazemitabaar2025exploring}. Contribution of AI should be staged and gated by quick checks~\citep{robe2022designing,kazemitabaar2025exploring}. Embodied variants can sustain engagement with the task when a studying partner is absent, as mixed reality pair-programming with a conversational avatar ties feedback to editor context~\citep{manfredi2023mixed}. Finally, to reduce over-reliance, systems should make performance on similar tasks completed without assistance visible for self-assessment~\citep{kazemitabaar2025exploring}.}

Beyond interaction paradigms, \change{17 out of 28} studies explore design principles for \ac{ai}-powered systems inside \acp{ide}, \change{for instance~\citep{chen2024design, robe2022designing, jayagopal2022exploring,cheng2024would,yen2023coladder,shlomov2024ida,guo2024investigating,mcnutt2023design,wang2023slide4n,feng2024coprompt,gu2024data,angert2023spellburst,kazemitabaar2024improving}}. These principles guide how \ac{ai} should be embedded into developer workflows and environments while maximizing usability, reliability, and efficiency:
\begin{itemize}
    \item \textbf{Context Awareness}:  LLM-powered assistance significantly boosts task completion rates when it can effectively incorporate a developer’s code as context~\citep{nam2023ide,gu2024data}.
    \item \textbf{Explainability \& Transparency}: Tools should provide context-aware explanations and highlight uncertain \ac{ai}-generated code~\citep{vasconcelos2025generation, yan2024ivie, sun2022investigating, wang2024investigating,vaithilingam2023towards} to improve user trust.
    \item \textbf{User Control \& Adaptability}: Users need adaptive, fine-tuned to expertise or business context \ac{ai} responses, and scaffolded \ac{ai} guidance to prevent over-reliance~\citep{prather2024widening, jiang2022discovering, cheng2024prompt}.
\end{itemize}
 
Among implementations of \ac{ai} for \acp{ide}, GitHub Copilot is by far the most studied, with \change{36 out of 90} works investigating it. This concentrated attention on a single tool suggests a risk of overgeneralization, where findings may not fully translate to other \ac{ai}-powered coding assistants, limiting broader design and implementation insights. 

However, research also explores alternative \ac{ai}-powered tools and approaches beyond Copilot, highlighting a growing landscape of intelligent development assistants. For example, IDA~\citep{shlomov2024ida} introduces a no-code automation system tailored for non-programmers, while Prompt Sapper~\citep{cheng2024prompt} supports \ac{ai} chain development through modular and visual programming interfaces. Studies, such as CoLadder~\citep{yen2023coladder}, investigate hierarchical prompt structures to refine interactions with code-generation models. PairBuddy~\citep{robe2022designing} explores conversational \ac{ai} as a simulated pair-programming partner. Domain-specific solutions, such as Slide4N~\citep{wang2023slide4n} for human-\ac{ai} collaboration in computational notebooks and RealHumanEval~\citep{mozannar2024realhumaneval} for LLM performance evaluation in software engineering, illustrate the diversity of \ac{ai} integrations in development workflows.

\subsection{Quality}

Investigating and ensuring the \textit{Quality} of \ac{ai}-assisted code generation is a central concern of \change{19 out of 90 papers in the reviewed literature (see Fig.~\ref{fig:scope} for IDs).} Across studies, quality is examined from multiple perspectives, including code correctness, security, and readability. 

One recurring theme is the trade-off between efficiency and correctness. \ac{ai}-assisted coding significantly accelerates development workflows, but at the cost of increased susceptibility to subtle errors, security vulnerabilities, and reduced maintainability~\citep{sandoval2023lost, tang2024study,pearce2022asleep}. Several studies suggest that while \ac{ai}-generated code often appears syntactically valid, it can contain logical flaws, insecure patterns, or deviations from best practices~\citep{asare2024user, pearce2022asleep}. As a result, effective quality control mechanisms, such as automated verification, static analysis, and model-aware debugging tools, are emphasized as critical to responsible adoption~\citep{agarwal2024copilot}. 

Beyond correctness, readability and maintainability emerge as key aspects of \ac{ai}-generated code quality. Although \ac{ai} can generate complex solutions rapidly, its output is not always optimized for human understanding. Studies highlight that \ac{ai}-generated code may be less readable and comprehensible due to overly concise structures, unconventional variable naming, or lack of meaningful comments~\citep{al2022readable}. This raises concerns about long-term maintainability, particularly in team-based software development workflows, where code legibility is as important as functional correctness. Research suggests that integrating \ac{ai}-powered refactoring and explanation tools can help mitigate these challenges, enhancing both short-term efficiency and long-term code quality~\citep{yan2024ivie}. 

Studies additionally pose security as an additional concern: up to 36\% of vulnerabilities in \ac{ai}-assisted code originate from the LLMs, pointing out the risks of replicating insecure patterns from training data~\citep{sandoval2023lost}, emphasizing the need for cautious oversight of \ac{ai} suggestions.

Overall, while LLM-powered tools can improve the software development process, maintaining consistently high-quality code remains a challenge. Ensuring reliability requires better alignment with developers’ needs and robust verification mechanisms~\citep{tang2024study, agarwal2024copilot}. \change{Studies highlight the need for strategies such as filtering suboptimal suggestions, improving contextual prompting through docstrings and function names~\citep{mozannar2024realhumaneval, de2024transformer}. Moreover, research calls for adaptive personalisation of code suggestions to support the overall quality of AI-generated code. In practice, this means learning per-developer and per-task policies for when to suggest, which modality to use, how much code to propose, and how much context to include in the prompt~\citep{mozannar2024realhumaneval, de2024transformer}}.

Quality of code in \ac{ai}-assisted development is not just a technical issue but also a human and organizational challenge. Research highlights that quality outcomes are influenced by how developers interact with \ac{ai} tools: whether they verify \ac{ai}-generated code, how they prompt \ac{ai} models, and whether they develop strategies for \ac{ai}-assisted debugging~\citep{weber2024significant, mozannar2024realhumaneval, sandoval2023lost, al2022readable, de2024transformer}. Best practices for improving \ac{ai}-assisted development quality extend beyond tool design and include education, training, and overall workflow adaptation.

\subsection{Methodologies}

There are multiple methodological approaches used in human-\ac{ai} interaction research. These methodologies can broadly be categorized into \emph{qualitative} and \emph{quantitative} approaches. Quantitative methods, in turn, can be further divided into survey-based studies and experimental research designs. Among 90 papers reviewed in this study, 12 employed a survey methodology, 32 followed an experimental design, and 38 utilized qualitative methods. 

Qualitative studies included a range of approaches, from semi-structured interviews to more complex methods such as the Wizard of Oz paradigm~\citep{robe2022designing}, focus group interviews~\citep{vaithilingam2023towards}, and Grounded Theory-based analysis~\citep{li2024ai}. Some studies adopted a mixed-methods approach, combining multiple methodologies. For instance, an experimental design complemented by participant interviews (a format often referred to as a user study). 

Additionally, 20 of the reviewed studies did not fit neatly into the qualitative-quantitative dichotomy. Examples include case studies in which the authors both describe their own experiences and provide quantitative metrics on system behavior~\citep{ross2023case}, quantitative analysis of system behavior using pre-defined datasets of typical tasks~\citep{nguyen2022empirical, fagadau2024analyzing}, and investigations of public discussions on platforms such as Reddit~\citep{klemmer2024using} and GitHub Issues \citep{zhou2025exploring}.

The sample sizes for different methodological approaches varied considerably. For interview-based studies, the sample size ranged from 7 to 61 participants (median = 16). Experimental designs involved sample sizes ranging from 17 to 214 participants, with the exception of two A/B studies, one with 535 participants~\citep{mozannar2024show} and another, conducted by Meta \citep{dunay2024multi}, reporting ``thousands'' of participants. As expected, survey-based studies had the largest sample sizes, ranging from 68 to 2,047 participants (median = 507).

\change{\subsection{Future Work Suggested by Existing Literature}

To address RQ3, we analyzed 250 future work statements extracted from the reviewed papers (see the full table in~\citep{hax_dataset_2025}). Each item was open-coded by topic (what is proposed) and method (how it is to be studied), resulting in 17 topics and 10 methodological strategies. 

\begin{figure*}[tb]
    \centering
    \includegraphics[width=.65\linewidth]{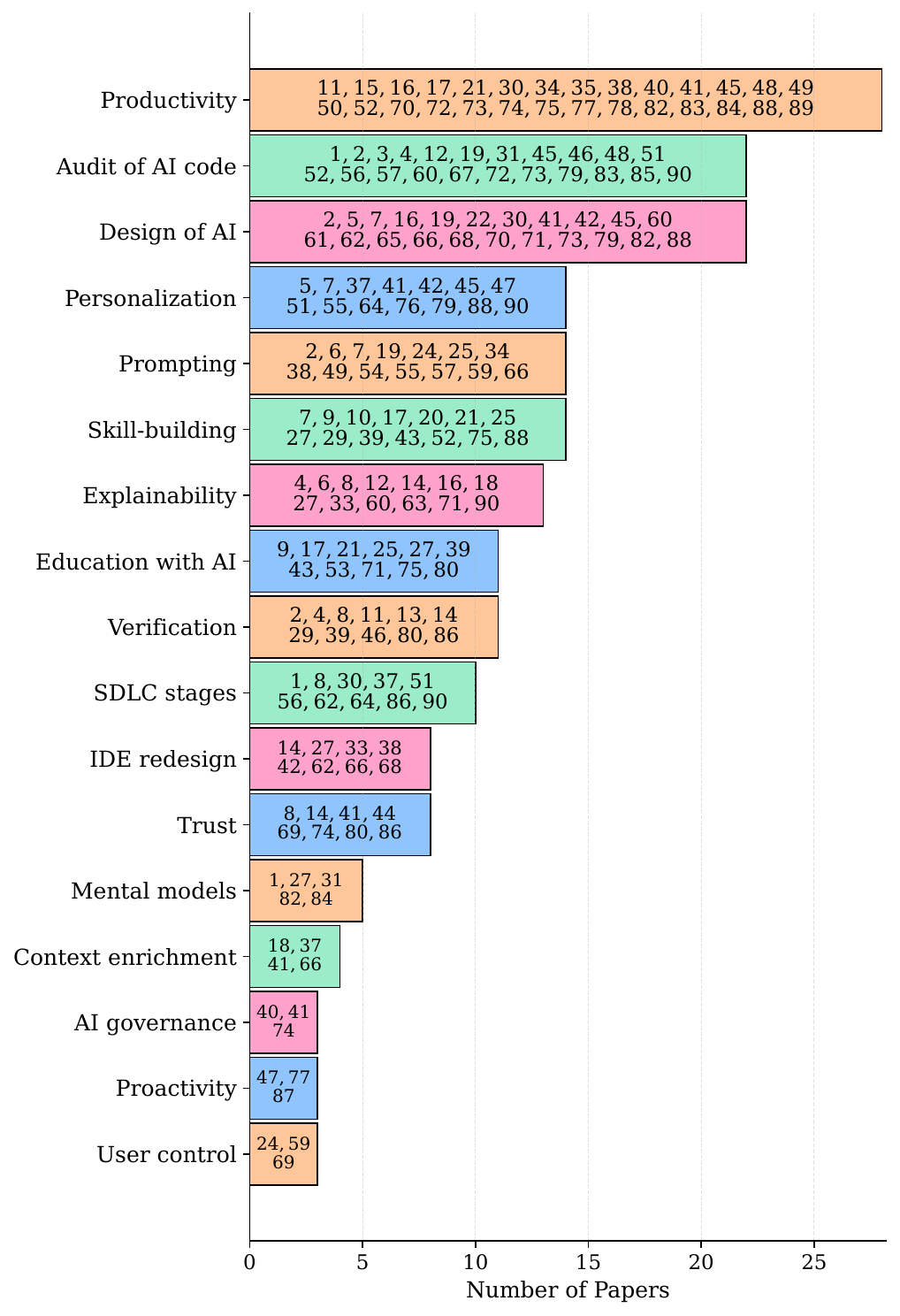}
    \caption{\change{Paper IDs for Topics of suggested future work.}}
    \label{fig:futurework_topics}
\end{figure*}

In Figure~\ref{fig:futurework_topics}, we present corresponding Paper IDs for each topic, and Figure~\ref{fig:futurework_heatmap} visualizes the distribution of papers across combinations between topics and methods. 

\begin{figure*}[tb]
    \centering
    \includegraphics[width=\linewidth]{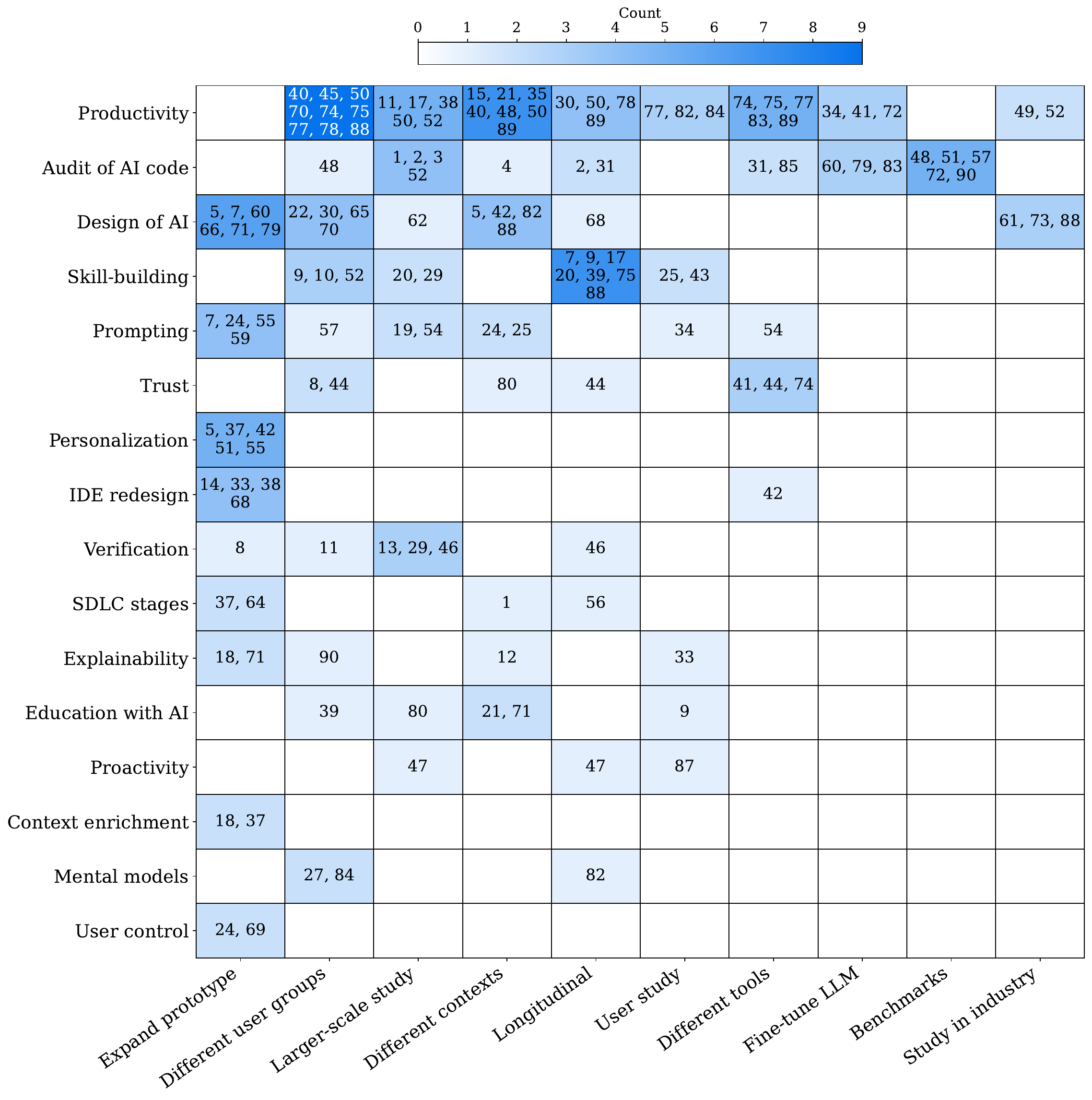}
    \caption{\change{Paper IDs for co-occurrence between \textsc{Topic} on the y-axis and \textsc{Method} on the x-axis for suggested future work.}}
    \label{fig:futurework_heatmap}
\end{figure*}

By analyzing the future work proposed by each study alongside the goals of others, we identified alignment patterns and gaps that highlight both the progress made and the opportunities missed. Some future work suggestions align with the goals of other studies, but these connections are not always explicit, reflecting a fragmented research landscape. Common topics like productivity, audit of AI-generated code, and usability of designed AI tools appear across studies but are addressed in isolated contexts without direct cross-referencing. This lack of continuity limits the field's ability to build cumulative knowledge effectively.

Suggestions cluster in a set of 17 themes. The most common address productivity factors (43), the design of AI assistance (29), and audits of AI-generated code (28). Additional areas include prompting support (19), skill building with AI (17), education with AI (16), personalization (16), explainability and transparency (15), verification support (13), and broader coverage of the SDLC (12). Less frequent but present are trust and reliability (11), IDE redesign for AI (10), assistant context enrichment (6), user mental models (5), proactivity (4), user control (3), and AI governance (3).

Productivity-oriented suggestions emphasize broader sampling and comparative designs, often at a larger scale~\citep{al2022readable,prather2023s,angert2023spellburst,jiang2024analysis,rasnayaka2024empirical}. They ask for studies with professional developers in realistic settings~\citep{peng2023impact,li2024ai,jiang2024analysis,rasnayaka2024empirical,zhu2024towards,kazemitabaar2025exploring} and for attention to students and novices~\citep{jiang2024analysis,amoozadeh2024trust,kazemitabaar2025exploring}. Authors also recommend evaluations that account for project and environment factors~\cite{peng2023impact,feldman2024non,zhou2025exploring}. Several papers connect productivity to interface and workflow choices and propose concrete design changes and controls~\citep{ziegler2022productivity, angert2023spellburst, rasnayaka2024empirical, sagodi2024methodology, pandey2024transforming, amoozadeh2024trust, nam2023ide, jiang2024analysis}.

Future work on auditing AI-generated code seeks wider security coverage, longitudinal observation, and stronger assets for evaluation. Proposals include expanding vulnerability categories and language coverage and tracking change over time~\citep{pearce2022asleep,asare2023github,asare2024user}, building targeted test suites and benchmarks~\citep{liu2024empirical}, and increasing unit tests and quality metrics to stress code quality~\citep{yetistiren2022assessing,liu2024empirical}. Some papers extend audits to SDLC touchpoints and to interfaces that surface risk at decision time~\citep{liang2024large, asare2024user, liu2024empirical, kruse2024can, pearce2022asleep, vasconcelos2025generation}.

Prompting support centers on guided interactions and safe practice. Suggestions include structured conversational prompting and refinement aids~\citep{jiang2022discovering, angert2023spellburst}, prompt engineering strategies with comparative evaluation~\citep{pearce2022asleep, denny2023conversing, mastropaolo2023robustness, fagadau2024analyzing}, multi-modal prompt channels for tasks that are not purely textual~\citep{angert2023spellburst}, and prompt sanitization with privacy-aware feedback~\citep{li2024ai}. Authors also ask for generalization across tools and settings~\citep{fagadau2024analyzing}.

Personalization proposals focus on adapting assistance to user profiles, expertise, and task state. Examples include user models to personalize initial prompts and responses~\citep{ross2023programmer, chopra2024exploring}, frequency and snooze controls to manage attention~\citep{vaithilingam2023towards}, adaptive strategies that respond to real time performance and needs~\citep{kazemitabaar2025exploring}, and results that better match the intended coding context and style~\citep{de2024transformer, liu2024empirical, sahoo2024ansible, cheng2024prompt, mozannar2024realhumaneval, vasconcelos2025generation}. Several papers request tailored documentation and scaffolds for specific roles such as data analysts~\citep{wang2022documentation, gu2024data}.

Explainability and transparency aim to help users understand, validate, and teach the assistant. Suggested mechanisms include rationales, annotations, and code level explanations~\citep{vaithilingam2022expectation, hu2022practitioners, mcnutt2023design}, interpretability features that support accurate mental models and calibrated expectations~\citep{weisz2022better, jiang2022discovering, sun2022investigating, bird2022taking}, internal deliberation to improve reasoning quality~\citep{ross2023case}, and presentation of validation evidence such as testing suite results or multi model checks~\citep{omidvar2024evaluating, yan2024ivie}.

Verification support concentrates on tools that help users check and repair outputs before adoption. Proposals include automatic test generation and retrieval of reference examples~\citep{vaithilingam2022expectation, ferdowsi2024validating}, post processing and filtering layers that repair or block risky outputs~\citep{pearce2022asleep, al2022readable, ferdowsi2024validating}, interface patterns that present multiple alternatives for comparison~\citep{weisz2022better}, and study designs that measure how learners verify AI suggestions and develop durable checking habits~\citep{kazemitabaar2023novices, kazemitabaar2023studying}.

Coverage beyond the implementation stage remains a consistent request. Papers propose research and tooling for requirements, design, evolution, debugging, testing, deployment, and code search across the life cycle~\citep{nguyen2022empirical, liu2024empirical, obrien2024prompt, chopra2024exploring, ferdowsi2024validating, vasconcelos2025generation}. Many of these suggestions pair life cycle scope with deeper integration into the developer workspace~\citep{chopra2024exploring, ferdowsi2024validating}.

Trust and reliability suggestions examine individual and community factors as well as interface effects. Authors ask for studies of how communities and organizations form and recalibrate trust over time~\citep{cheng2024would, prather2024widening}, for quantification of interface-driven trust shifts~\citep{wang2024investigating}, for evaluations that measure the impact of assistant behavior changes on acceptance~\citep{feldman2024non}, and for approaches that foster appropriate trust with clear boundaries~\citep{ferdowsi2024validating}. Inclusion of diverse developer groups appears throughout~\citep{vaithilingam2022expectation, hu2022practitioners, ross2023programmer}.

Proposals on IDE redesign and context enrichment argue for AI-aware environments. Suggestions include non-linear input and new interaction media~\citep{mcnutt2023design}, improved error handling and better merging of sketch-like artifacts~\citep{angert2023spellburst}, chat or dialogue interfaces for automation control~\citep{shlomov2024ida}, and consistent propagation of edits and assumptions~\citep{kazemitabaar2024improving}. For context, papers recommend memory management, dynamic prompts that reflect project artifacts, and search-based integrations~\citep{ross2023case, wang2023slide4n, ross2023programmer, shlomov2024ida}.

Less frequent but important themes include user mental models, proactivity, user control, and governance. Future work asks for theory building on how developers form and use mental models of assistants and for longitudinal observation of expectation change~\citep{nguyen2022empirical, lau2023ban, asare2023github, zhu2024towards, amoozadeh2024trust}. Proactivity-related suggestions seek predictive models of user state and careful study of long-term effects on quality of code and productivity~\citep{de2024transformer, mozannar2024reading, mozannar2024show}. Use control-oriented suggestions call for clearer and richer controls over generation and automation levels~\citep{yen2023coladder, feng2024coprompt, wang2024investigating}. Governance suggestions raise questions about fairness, access, and alignment, including alignment toward helpful and harmless behavior and possible career impacts for different demographic groups~\citep{ross2023programmer, feldman2024non, peng2023impact}.

Taken together, these directions indicate the need for larger and longer evaluations, stronger audit and verification assets, broader life cycle coverage, and adaptive assistance that is explainable and under user control.
}

\section{Discussion}

\change{We organize the discussion of our systematic literature review of 90 in-IDE HAX studies around topics coverage and gaps (RQ1), implications for practice and research (RQ2), and a future work agenda (RQ3).}

\change{\paragraph{\textbf{RQ1: Extensively studied and under-explored aspects of in-IDE HAX.}}
The corpus of reviewed works separates cleanly into two contexts of use for \ac{ai} in software development, namely professional (71/90) and educational (19/90). Within professional settings, most papers examine practical use in coding and maintenance, and a large share evaluate assistance without situating it in a specific \ac{sdlc} stage. Educational papers focus on how learners use assistance to solve programming problems. Here as well, stage information is often omitted, which limits comparison across contexts. This weakens interpretability because the same interface may have different effects at requirements, implementation, testing, or evolution. Future work should report the stage explicitly and use a simple task taxonomy that allows cross-paper synthesis. Task design employed in studies evolves over time. Early studies rely on small algorithmic exercises that resemble simple games and interview-style problems. More recent work introduces project-level analysis of workflows and begins to include tasks related to documentation, testing, debugging, and migration. This broadening improves ecological validity, yet it also increases the number of papers that label the task as unspecified. Clearer task descriptions and stable taxonomies are needed to preserve comparability as the scope widens.

Several papers frame ~\ac{ai}-assistance as a form of pair programming in which the system acts as a collaborator rather than a tool. This view aligns with the observed move from isolated completions to co-creative workflows. It also raises concrete needs in both contexts, namely training in model interaction and oversight in professional teams, and scaffolds that keep learners in control in educational settings. In summary, the evidence is richest for professional coding and maintenance, with a growing variety in task types and a notable reporting gap on the \ac{sdlc} stage. The next step is to include non-adopters, expand coverage to early and late stages, and standardize task and stage reporting so that results can be compared across settings and over time.}

\change{\paragraph{\textbf{RQ2: Key findings and implications in the field of HAX in IDEs.\newline}}

\noindent\hspace*{\parindent}\emph{Impact.}
Across reviewed studies, there is evidence that AI assistance in \acp{ide} changes how developers create and verify code. The main benefit is faster progress on routine or well-scoped work, often explained by fewer context switches and reduced need for boilerplate code writing. The main limit appears on tasks that embed proprietary knowledge or complex logic, where current systems struggle to capture project-specific nuance~\citep{peng2023impact, pandey2024transforming, khemka2024toward}. Code verification is, therefore, central. About fifty percent of developers' time is now spent on inspecting suggestions, refining prompts, and repairing code~\citep{mozannar2024reading}. These findings argue for budgeting verification explicitly and for evaluating impact with edits and verification effort in addition to the suggestions' acceptance rate and time spent on task. Effects are heterogeneous across populations. Novices complete tasks more quickly and report lower workload when the use of in-IDE AI is structured, yet they are vulnerable to subtle errors and may overtrust suggestions~\citep{kazemitabaar2023studying, prather2024widening, zhou2023case}. Professionals benefit from speed on familiar patterns but may under-rely on AI and reject correct suggestions if it lacks a clear rationale~\citep{sun2022investigating}. Trust also varies by context. It decreases in high-stakes and testing contexts and increases for routine or proof-of-concept work, which argues for visible boundaries and rationale at decision time~\citep{wang2024investigating, brown2024identifying}.

\emph{Design.}
Reviewed studies converge on three core design principles for \ac{ai} in \acp{ide}, namely rich context awareness, explainability with transparency, and user control with adaptability. These principles guide how AI systems should be built. Moreover, it is highlighted that the two dominant interaction modes, autocompletion and conversational, should work together. A smooth switch between them, a shared context pool, and consistent state across views allow developers to move from local edits to higher-level reasoning without friction~\citep{weber2024significant, yen2023coladder}. As conversational interaction with AI enters IDEs, prompting becomes an important learnable activity. Structured prompts and reusable prompt-outcome pairs align generation with intent and support calibrated acceptance~\citep{kruse2024can, denny2023conversing, fagadau2024analyzing}. Interfaces that ask the user to state the next step before code is shown and that supply brief rationales further stabilize decisions~\citep{kazemitabaar2025exploring}, as well as step-by-step explanations of the generated code for code validation and refinement~\cite{tian2023interactive,tian2024sqlucid}. Explanation and control must be available at decision time. Useful aids in verification include surfacing runtime values, side-by-side alternatives, retrieval of reference examples, static analysis, and unit tests inside the editor~\cite{ferdowsi2024validating,pearce2022asleep,weisz2022better,vaithilingam2022expectation}. Context-aware explanations and explicit scope controls make it clear where and why the assistant acts~\citep{vaithilingam2023towards, wang2024investigating, yan2024ivie}. When available, calibrated uncertainty cues or predicted edit likelihood should direct review effort, without relying on raw generation probabilities alone~\citep{spiess2025calibration,vasconcelos2025generation}. In educational settings, structured prompts, brief rationales, staged contribution, and quick checks inside the \ac{ide} help learners stay in control~\citep{robe2022designing, kazemitabaar2024codeaid, kazemitabaar2023studying, prather2024widening, zhou2023case}. Embodied or mixed reality agents are useful when the goal is to practice collaboration routines with feedback tied to the editor context~\citep{manfredi2023mixed}.

\emph{Quality.}
Studies in this direction examine correctness, security, readability, and maintainability of code produced with AI assistance inside the \ac{ide}. Across these papers, the central pattern is a trade-off between speed and assurance. Assistance can accelerate progress, yet LLM-generated code that passes superficial checks may still contain logical flaws or insecure patterns, which shifts effort from authoring toward checking \citep{wang2025towards,nguyen2022empirical,yetistiren2022assessing,al2022readable,asare2023github,sandoval2023lost,asare2024user,mozannar2024reading,izadi2024language}. LLM as judge signals can complement human review for correctness~\citep{tong2024codejudge}, but correctness outcomes depend not only on model properties but also on developer practice, including how prompts are written, how suggestions are inspected, and how teams organize review \citep{kruse2024can,feng2024coprompt,kuang2024developers,mozannar2024reading}. Readability and maintainability concerns recur alongside correctness. Short and compact suggestions reduce typing but can hinder clarity of intent through unconventional names, missing comments, or terse structure, which complicates later edits and team review \citep{nguyen2022empirical,yetistiren2022assessing,vaithilingam2022expectation,al2022readable}. Explanatory aids and refactoring support inside the editor help address these issues and make suggested changes easier to justify \citep{yan2024ivie,ferdowsi2024validating}. Security is a shared responsibility as well. Models can reproduce unsafe templates from training data, while developers may accept risky code under time pressure \citep{pearce2022asleep,sandoval2023lost,asare2023github,asare2024user,al2022readable,mozannar2024reading,vaithilingam2022expectation,gardella2024performance}. Taken together, these findings call again for verification with tests and static analysis inside the editor, retrieval of relevant examples, and concise explanations at decision time, so that speed and assurance can coexist \citep{vaithilingam2022expectation,liu2024empirical,yan2024ivie,ferdowsi2024validating}.}

\paragraph{\textbf{RQ3: Future research and development agenda.\newline}}
\change{Analysis of the future work statements yields a ranked set of topics and methodologies to study in-IDE HAX. The most frequent cluster concerns productivity factors \((43)\). Here, the field should move from isolated gains to comparative and larger-scale evaluations that report acceptance, edits, and verification effort in addition to time. The next clusters focus on the design of assistance \((29)\) and audits of \ac{ai} generated code \((28)\). For design, work should advance hybrid workflows that integrate autocompletion and conversation, with prompt scaffolds and explanation layers tied to decision points \citep{yen2023coladder,ross2023programmer,brown2024identifying,guo2024investigating}. For audit, the field needs targeted test suites, stronger security coverage, and longitudinal assets that track change over time \citep{agarwal2024copilot,mozannar2024realhumaneval,asare2024user,pearce2022asleep,asare2023github,sagodi2024methodology}. Prompting support appears in nineteen statements and should prioritize guided decomposition and hierarchical specification, with interactive explanations that support validation and refinement \citep{yen2023coladder,brown2024identifying}. Skill building with \ac{ai} \((17)\) and education with \ac{ai} \((16)\) call for structured practice that protects learning goals inside authentic \ac{ide} settings \citep{prather2023s,zhou2023case,angert2023spellburst,gardella2024performance,ferdowsi2024validating}. Personalization is equally frequent \((16)\) and should aim at policies that adapt timing, modality, scope, and context to user profile and task state, with simple controls for frequency and snooze \citep{vaithilingam2023towards,ross2023programmer,de2024transformer,chopra2024exploring, koohestani2025rethinking}. Explainability and transparency are requested by fifteen papers and should be evaluated for their effect on acceptance. Verification support \((13)\) and broader \ac{sdlc} coverage \((12)\) align with the observed gaps in stage specification and with the shift of time toward checking. Less frequent but still important topics for study are trust and reliability \((11)\), \ac{ide} redesign for \ac{ai} \((10)\), assistant context enrichment \((6)\), user mental models, proactivity \((4)\), user control \((3)\), and governance \((3)\). These numbers do not imply that lower frequency topics are unimportant. They indicate where evidence and assets are thin and where careful study can yield high value.
}

\paragraph{\textbf{Methodological and open science recommendations.}}
\change{Across the corpus, sample sizes are often underpowered and rarely justified. The median is seventeen participants. According to prior studies, for HCI practice, the most common sample size is twelve and the median is eighteen; in-person studies have a median of fifteen, while remote studies have a median of seventy-seven~\citep{caine2016local}. Although many quantitative studies at these sizes will be underpowered and therefore benefit from replication. We recommend a priori power analysis for confirmatory comparisons and saturation arguments for qualitative designs, with explicit reporting of recruitment constraints. Collaboration between academia and industry can improve recruitment and enable longitudinal assets. To support replication, we recommend sharing versioned repositories with study materials and data where feasible, including codebooks, prompts, task descriptions, analysis scripts, and deidentified datasets with clear metadata and a data dictionary. We also recommend using preregistration, blinding, and randomization when appropriate, and embedding transparency checklists that record methodology, exclusion criteria, and data sharing plans.}

\change{\paragraph{\textbf{Takeaways.}}
Treat verification as a main part of the interaction with \ac{ai} assistants, keeping tests, static analysis, and reference examples inside the editor and by surfacing runtime evidence to support calibrated acceptance decisions. Evaluate the effect beyond time spent on the task by reporting acceptance, edits, and verification effort. Design \ac{ai} assistance as a hybrid of autocompletion and conversation that shares context, offers brief explanations, and scope controls at decision time. Structure educational use of \ac{ai} in IDEs with staged contribution of \ac{ai}, stepwise reveal of its suggestions, and controls for their quick checks. For studies, report the \ac{sdlc} stage and expand coverage of tasks completed with \ac{ai} to early and late stages. Strengthen methodological rigor through justified sample sizes, preregistration where appropriate, transparent reporting, and collaboration with industry for larger, ecologically valid, and longitudinal evaluations.}

\subsection{Threats to the Validity}
While gathered findings provide valuable insights, in any scientific study, it is essential to consider potential threats to validity, which are factors that can affect the accuracy and generalizability of the research findings. 

\textbf{Sampling Bias:} \change{Despite efforts to include well-known libraries and refine the search string, the possibility of sampling bias remains. Our recall-oriented query design can bias the pool toward professional contexts, since educational deployments do not always occur inside full IDEs or are indexed under alternative platforms. We mitigated this by including multiple synonyms where supported and by manually screening every record for the inclusion criterion. Although our approach aimed to minimize bias, the inherent challenge of capturing every relevant work persists. That is why we provide our search protocol to mitigate this threat.}

\textbf{Temporal Bias:} The chosen time frame (papers published between 2022 and 2024) introduces a potential temporal bias, excluding earlier works. This decision was driven by the intention to focus on contemporary developments following the advent of LLMs. While acknowledging potential temporal limitations, we aim to capture the latest advancements and trends in this rapidly evolving field.

\textbf{Source Reliability:} Inclusion of non-peer-reviewed papers from ArXiv introduces concerns regarding the reliability of findings, given the absence of a formal peer-review process. Recognizing this limitation, we deemed it necessary to consider insights from ArXiv due to the dynamic and rapidly evolving nature of the field. Thoroughly examining titles, abstracts, and full texts ensured that all included works contributed meaningfully to our survey. Moreover, the open publication environment of ArXiv encourages the publication of negative or null results, contributing to a more balanced representation of the research outcomes.

\textbf{Interpretation Bias:} Analyzing a large amount of information can introduce interpretation bias and impact the way studies are categorized. While acknowledging this complexity, we emphasize transparency and provide the entire dataset~\citep{hax_dataset_2025} for the readers. 

\textbf{Challenges of industry-related research:} It is important to acknowledge that, overall, in-IDE \ac{hax} research faces unique challenges due to its close ties with the industry and professional context. One of the most significant challenges of conducting empirical studies in industrial settings is the "time factor". Industrial sponsors and participants often have short time horizons. 
They expect valuable feedback within 1-2 months of research, so that they could make timely business decisions. Although industrial sponsors and project managers are interested in rigorous results, practical constraints sometimes make it challenging to prioritize statistical significance. Analyzing immediate trends and patterns can be sufficient for them to assess the performance of their projects, teams, or technologies.
By fostering a collaborative environment, we can develop strategies that balance the demands of industrial collaboration with the standards of academic research.

\section{CONCLUSION}

\change{This review offers a structured synthesis of current research on AI-powered assistance in integrated development environments. Drawing on 90 studies, we identified dominant trends, conceptual gaps, and methodological characteristics of the emerging field of \ac{in-ide-hax}.

We found that research to date focuses heavily on a small set of tools, most often GitHub Copilot, and concentrates primarily on the implementation stage of the \ac{sdlc}. While this has produced early insights into how developers engage with \ac{ai} assistance during code writing and modification, it leaves other critical stages such as requirements analysis, testing, and deployment underexplored. To make progress, the field must broaden its lens. Future research should investigate diverse \ac{ai} assistants across varied development environments, extend coverage to earlier and later lifecycle stages, and prioritize longitudinal and comparative designs that capture how practices and their impact evolve over time. Attention is also needed to support personalization and trust calibration, reduce validation burden, and address broader quality and governance concerns. Finally, we note that methodological rigor remains an area for improvement. Methodologically, studies tend to prioritize short-term evaluations, often with underpowered samples and limited contextual variation. As the field matures, advancing cumulative evidence will require the adoption of robust empirical practices and closer collaboration across research groups.

In sum, the reviewed literature demonstrates growing interest in the integration of \ac{ai} into developer tools but also highlights the need for deeper empirical grounding, broader tool coverage, and stronger methodological foundations. Addressing these challenges will be essential for supporting effective, transparent, and user-centered \ac{ai} assistance in software development workflows.}

\begin{acknowledgements}
This work was conducted as part of the AI for Software Engineering (AI4SE) collaboration between JetBrains and Delft University of Technology. The authors gratefully acknowledge the financial support provided by JetBrains, which made this research possible.
\end{acknowledgements}

\section*{Declarations}
\subsection*{Funding}
This research was supported by JetBrains Research.

\subsection*{Ethical approval}
Not applicable.

\subsection*{Informed consent}
Not applicable.

\subsection*{Author Contributions}
\noindent Agnia Sergeyuk, Ilya Zakharov, Ekaterina Koshchenko: Conceptualization, methodology, literature survey, writing, original draft, and editing.  

\noindent Maliheh Izadi: Conceptualization, methodology, writing, original draft, and review.

\subsection*{Data availability}
All data are available in our supplementary materials~\citep{hax_dataset_2025}.

\subsection*{Conflict of interest}
The authors declare that they have no conflict of interest.

\subsection*{Clinical Trial Number} 
Not applicable.

\bibliographystyle{spbasic}
\bibliography{main_refs}

\begin{thebibliography}{117}
\providecommand{\natexlab}[1]{#1}
\providecommand{\url}[1]{{#1}}
\providecommand{\urlprefix}{URL }
\expandafter\ifx\csname urlstyle\endcsname\relax
  \providecommand{\doi}[1]{DOI~\discretionary{}{}{}#1}\else
  \providecommand{\doi}{DOI~\discretionary{}{}{}\begingroup \urlstyle{rm}\Url}\fi
\providecommand{\eprint}[2][]{\url{#2}}

\bibitem[{Agarwal et~al.(2024)Agarwal, Chan, Chandel, Jang, Miller, Moghaddam, Mohylevskyy, Sundaresan, and Tufano}]{agarwal2024copilot}
Agarwal A, Chan A, Chandel S, Jang J, Miller S, Moghaddam RZ, Mohylevskyy Y, Sundaresan N, Tufano M (2024) Copilot evaluation harness: Evaluating llm-guided software programming. arXiv preprint arXiv:240214261

\bibitem[{Agbo et~al.(2025)Agbo, Olivia, Oguibe, Sanusi, and Sani}]{agbo2025computing}
Agbo FJ, Olivia C, Oguibe G, Sanusi IT, Sani G (2025) Computing education using generative artificial intelligence tools: A systematic literature review. Computers and Education Open p 100266

\bibitem[{Al~Madi(2022)}]{al2022readable}
Al~Madi N (2022) How readable is model-generated code? examining readability and visual inspection of github copilot. In: Proceedings of the 37th IEEE/ACM international conference on automated software engineering, pp 1--5

\bibitem[{Alshamrani and Bahattab(2015)}]{alshamrani2015comparison}
Alshamrani A, Bahattab A (2015) A comparison between three sdlc models waterfall model, spiral model, and incremental/iterative model. International Journal of Computer Science Issues (IJCSI) 12(1):106

\bibitem[{Amoozadeh et~al.(2024)Amoozadeh, Daniels, Nam, Kumar, Chen, Hilton, Srinivasa~Ragavan, and Alipour}]{amoozadeh2024trust}
Amoozadeh M, Daniels D, Nam D, Kumar A, Chen S, Hilton M, Srinivasa~Ragavan S, Alipour MA (2024) Trust in generative ai among students: An exploratory study. In: Proceedings of the 55th ACM Technical Symposium on Computer Science Education V. 1, pp 67--73

\bibitem[{Angert et~al.(2023)Angert, Suzara, Han, Pondoc, and Subramonyam}]{angert2023spellburst}
Angert T, Suzara M, Han J, Pondoc C, Subramonyam H (2023) Spellburst: A node-based interface for exploratory creative coding with natural language prompts. In: Proceedings of the 36th Annual ACM Symposium on User Interface Software and Technology, pp 1--22

\bibitem[{Ariza et~al.(2025)Ariza, Restrepo, and Hern{\'a}ndez}]{ariza2025generative}
Ariza J{\'A}, Restrepo MB, Hern{\'a}ndez CH (2025) Generative ai in engineering and computing education: A scoping review of empirical studies and educational practices. IEEE Access

\bibitem[{Asare et~al.(2023)Asare, Nagappan, and Asokan}]{asare2023github}
Asare O, Nagappan M, Asokan N (2023) Is github’s copilot as bad as humans at introducing vulnerabilities in code? Empirical Software Engineering 28(6):129

\bibitem[{Asare et~al.(2024)Asare, Nagappan, and Asokan}]{asare2024user}
Asare O, Nagappan M, Asokan N (2024) A user-centered security evaluation of copilot. In: Proceedings of the IEEE/ACM 46th International Conference on Software Engineering, pp 1--11

\bibitem[{Barke et~al.(2023)Barke, James, and Polikarpova}]{barke2023grounded}
Barke S, James MB, Polikarpova N (2023) Grounded copilot: How programmers interact with code-generating models. Proceedings of the ACM on Programming Languages 7(OOPSLA1):85--111

\bibitem[{Bird et~al.(2022)Bird, Ford, Zimmermann, Forsgren, Kalliamvakou, Lowdermilk, and Gazit}]{bird2022taking}
Bird C, Ford D, Zimmermann T, Forsgren N, Kalliamvakou E, Lowdermilk T, Gazit I (2022) Taking flight with copilot: Early insights and opportunities of ai-powered pair-programming tools. Queue 20(6):35--57

\bibitem[{Brown et~al.(2024)Brown, D'Angelo, Murillo, Jaspan, and Green}]{brown2024identifying}
Brown A, D'Angelo S, Murillo A, Jaspan C, Green C (2024) Identifying the factors that influence trust in ai code completion. In: Proceedings of the 1st ACM International Conference on AI-Powered Software, pp 1--9

\bibitem[{Caine(2016)}]{caine2016local}
Caine K (2016) Local standards for sample size at chi. In: Proceedings of the 2016 CHI conference on human factors in computing systems, pp 981--992

\bibitem[{Carrera-Rivera et~al.(2022)Carrera-Rivera, Ochoa, Larrinaga, and Lasa}]{carrera2022conduct}
Carrera-Rivera A, Ochoa W, Larrinaga F, Lasa G (2022) How-to conduct a systematic literature review: A quick guide for computer science research. MethodsX 9:101895

\bibitem[{{Centre for Reviews and Dissemination (UK)}(1995)}]{DARE1995}
{Centre for Reviews and Dissemination (UK)} (1995) Database of abstracts of reviews of effects (dare): Quality-assessed reviews. \urlprefix\url{https://www.ncbi.nlm.nih.gov/books/NBK285222/}, accessed: 2025-08-10

\bibitem[{Chen and Zacharias(2024)}]{chen2024design}
Chen J, Zacharias J (2024) Design principles for collaborative generative ai systems in software development. In: International Conference on Design Science Research in Information Systems and Technology, Springer, pp 341--354

\bibitem[{Cheng et~al.(2024{\natexlab{a}})Cheng, Wang, Zimmermann, and Ford}]{cheng2024would}
Cheng R, Wang R, Zimmermann T, Ford D (2024{\natexlab{a}}) “it would work for me too”: How online communities shape software developers’ trust in ai-powered code generation tools. ACM Transactions on Interactive Intelligent Systems 14(2):1--39

\bibitem[{Cheng et~al.(2024{\natexlab{b}})Cheng, Chen, Huang, Xing, Xu, and Lu}]{cheng2024prompt}
Cheng Y, Chen J, Huang Q, Xing Z, Xu X, Lu Q (2024{\natexlab{b}}) Prompt sapper: A llm-empowered production tool for building ai chains. ACM Transactions on Software Engineering and Methodology 33(5):1--24

\bibitem[{Chopra et~al.(2024)Chopra, Bajpai, Biyani, Soares, Radhakrishna, Parnin, and Gulwani}]{chopra2024exploring}
Chopra B, Bajpai Y, Biyani P, Soares G, Radhakrishna A, Parnin C, Gulwani S (2024) Exploring interaction patterns for debugging: Enhancing conversational capabilities of ai-assistants. arXiv preprint arXiv:240206229

\bibitem[{Denny et~al.(2023)Denny, Kumar, and Giacaman}]{denny2023conversing}
Denny P, Kumar V, Giacaman N (2023) Conversing with copilot: Exploring prompt engineering for solving cs1 problems using natural language. In: Proceedings of the 54th ACM technical symposium on computer science education V. 1, pp 1136--1142

\bibitem[{Dunay et~al.(2024)Dunay, Cheng, Tait, Thakkar, Rigby, Chiu, Ahmad, Ganesan, Maddila, Murali et~al.}]{dunay2024multi}
Dunay O, Cheng D, Tait A, Thakkar P, Rigby PC, Chiu A, Ahmad I, Ganesan A, Maddila C, Murali V, et~al. (2024) Multi-line ai-assisted code authoring. In: Companion Proceedings of the 32nd ACM International Conference on the Foundations of Software Engineering, pp 150--160

\bibitem[{Durrani et~al.(2024)Durrani, Akpinar, Adak, Kabakus, Ozturk, and Saleh}]{durrani2024decade}
Durrani UK, Akpinar M, Adak MF, Kabakus AT, Ozturk MM, Saleh M (2024) A decade of progress: A systematic literature review on the integration of ai in software engineering phases and activities (2013-2023). IEEE Access

\bibitem[{Fagadau et~al.(2024)Fagadau, Mariani, Micucci, and Riganelli}]{fagadau2024analyzing}
Fagadau ID, Mariani L, Micucci D, Riganelli O (2024) Analyzing prompt influence on automated method generation: An empirical study with copilot. In: Proceedings of the 32nd IEEE/ACM International Conference on Program Comprehension, pp 24--34

\bibitem[{Feldman and Anderson(2024)}]{feldman2024non}
Feldman MQ, Anderson CJ (2024) Non-expert programmers in the generative ai future. In: Proceedings of the 3rd Annual Meeting of the Symposium on Human-Computer Interaction for Work, pp 1--19

\bibitem[{Feng et~al.(2024)Feng, Yen, You, Fan, Zhao, and Lu}]{feng2024coprompt}
Feng L, Yen R, You Y, Fan M, Zhao J, Lu Z (2024) Coprompt: Supporting prompt sharing and referring in collaborative natural language programming. In: Proceedings of the 2024 CHI Conference on Human Factors in Computing Systems, pp 1--21

\bibitem[{Ferdowsi et~al.(2024)Ferdowsi, Huang, James, Polikarpova, and Lerner}]{ferdowsi2024validating}
Ferdowsi K, Huang R, James MB, Polikarpova N, Lerner S (2024) Validating ai-generated code with live programming. In: Proceedings of the 2024 CHI Conference on Human Factors in Computing Systems, pp 1--8

\bibitem[{Gardella et~al.(2024)Gardella, Pettit, and Riggs}]{gardella2024performance}
Gardella N, Pettit R, Riggs SL (2024) Performance, workload, emotion, and self-efficacy of novice programmers using ai code generation. In: Proceedings of the 2024 on Innovation and Technology in Computer Science Education V. 1, Association for Computing Machinery, pp 290--296

\bibitem[{Gu et~al.(2024)Gu, Grunde-McLaughlin, McNutt, Heer, and Althoff}]{gu2024data}
Gu K, Grunde-McLaughlin M, McNutt A, Heer J, Althoff T (2024) How do data analysts respond to ai assistance? a wizard-of-oz study. In: Proceedings of the 2024 CHI Conference on Human Factors in Computing Systems, pp 1--22

\bibitem[{Guo et~al.(2024)Guo, Mohanty, Piazentin~Ono, Hao, Gou, and Ren}]{guo2024investigating}
Guo J, Mohanty V, Piazentin~Ono JH, Hao H, Gou L, Ren L (2024) Investigating interaction modes and user agency in human-llm collaboration for domain-specific data analysis. In: Extended Abstracts of the CHI Conference on Human Factors in Computing Systems, pp 1--9

\bibitem[{He et~al.(2025)He, Treude, and Lo}]{he2025llm}
He J, Treude C, Lo D (2025) Llm-based multi-agent systems for software engineering: Literature review, vision, and the road ahead. ACM Transactions on Software Engineering and Methodology 34(5):1--30

\bibitem[{Hou et~al.(2024)Hou, Zhao, Liu, Yang, Wang, Li, Luo, Lo, Grundy, and Wang}]{hou2024large}
Hou X, Zhao Y, Liu Y, Yang Z, Wang K, Li L, Luo X, Lo D, Grundy J, Wang H (2024) Large language models for software engineering: A systematic literature review. ACM Transactions on Software Engineering and Methodology 33(8):1--79

\bibitem[{Hu et~al.(2022)Hu, Xia, Lo, Wan, Chen, and Zimmermann}]{hu2022practitioners}
Hu X, Xia X, Lo D, Wan Z, Chen Q, Zimmermann T (2022) Practitioners' expectations on automated code comment generation. In: Proceedings of the 44th international conference on software engineering, pp 1693--1705

\bibitem[{Imai(2022)}]{imai2022github}
Imai S (2022) Is github copilot a substitute for human pair-programming? an empirical study. In: Proceedings of the ACM/IEEE 44th International Conference on Software Engineering: Companion Proceedings, pp 319--321

\bibitem[{Izadi et~al.(2024)Izadi, Katzy, Van~Dam, Otten, Popescu, and Van~Deursen}]{izadi2024language}
Izadi M, Katzy J, Van~Dam T, Otten M, Popescu RM, Van~Deursen A (2024) Language models for code completion: A practical evaluation. In: Proceedings of the IEEE/ACM 46th International Conference on Software Engineering, pp 1--13

\bibitem[{Jayagopal et~al.(2022)Jayagopal, Lubin, and Chasins}]{jayagopal2022exploring}
Jayagopal D, Lubin J, Chasins SE (2022) Exploring the learnability of program synthesizers by novice programmers. In: Proceedings of the 35th Annual ACM Symposium on User Interface Software and Technology, pp 1--15

\bibitem[{{JetBrains}(2024)}]{JetBrains2024AI}
{JetBrains} (2024) {The State of Developer Ecosystem 2024: AI Insights}. \urlprefix\url{https://www.jetbrains.com/lp/devecosystem-2024/#ai}, accessed: 2025-08-10

\bibitem[{Jiang et~al.(2022)Jiang, Toh, Molina, Olson, Kayacik, Donsbach, Cai, and Terry}]{jiang2022discovering}
Jiang E, Toh E, Molina A, Olson K, Kayacik C, Donsbach A, Cai CJ, Terry M (2022) Discovering the syntax and strategies of natural language programming with generative language models. In: Proceedings of the 2022 CHI Conference on Human Factors in Computing Systems, pp 1--19

\bibitem[{Jiang and Coblenz(2024)}]{jiang2024analysis}
Jiang S, Coblenz M (2024) An analysis of the costs and benefits of autocomplete in ides. Proceedings of the ACM on Software Engineering 1(FSE):1284--1306

\bibitem[{Kazemitabaar et~al.(2023{\natexlab{a}})Kazemitabaar, Chow, Ma, Ericson, Weintrop, and Grossman}]{kazemitabaar2023studying}
Kazemitabaar M, Chow J, Ma CKT, Ericson BJ, Weintrop D, Grossman T (2023{\natexlab{a}}) Studying the effect of ai code generators on supporting novice learners in introductory programming. In: Proceedings of the 2023 CHI conference on human factors in computing systems, pp 1--23

\bibitem[{Kazemitabaar et~al.(2023{\natexlab{b}})Kazemitabaar, Hou, Henley, Ericson, Weintrop, and Grossman}]{kazemitabaar2023novices}
Kazemitabaar M, Hou X, Henley A, Ericson BJ, Weintrop D, Grossman T (2023{\natexlab{b}}) How novices use llm-based code generators to solve cs1 coding tasks in a self-paced learning environment. In: Proceedings of the 23rd Koli calling international conference on computing education research, pp 1--12

\bibitem[{Kazemitabaar et~al.(2024{\natexlab{a}})Kazemitabaar, Williams, Drosos, Grossman, Henley, Negreanu, and Sarkar}]{kazemitabaar2024improving}
Kazemitabaar M, Williams J, Drosos I, Grossman T, Henley AZ, Negreanu C, Sarkar A (2024{\natexlab{a}}) Improving steering and verification in ai-assisted data analysis with interactive task decomposition. In: Proceedings of the 37th Annual ACM Symposium on User Interface Software and Technology, pp 1--19

\bibitem[{Kazemitabaar et~al.(2024{\natexlab{b}})Kazemitabaar, Ye, Wang, Henley, Denny, Craig, and Grossman}]{kazemitabaar2024codeaid}
Kazemitabaar M, Ye R, Wang X, Henley AZ, Denny P, Craig M, Grossman T (2024{\natexlab{b}}) Codeaid: Evaluating a classroom deployment of an llm-based programming assistant that balances student and educator needs. In: Proceedings of the 2024 chi conference on human factors in computing systems, pp 1--20

\bibitem[{Kazemitabaar et~al.(2025)Kazemitabaar, Huang, Suh, Henley, and Grossman}]{kazemitabaar2025exploring}
Kazemitabaar M, Huang O, Suh S, Henley AZ, Grossman T (2025) Exploring the design space of cognitive engagement techniques with ai-generated code for enhanced learning. In: Proceedings of the 30th International Conference on Intelligent User Interfaces, pp 695--714

\bibitem[{Khemka and Houck(2024)}]{khemka2024toward}
Khemka M, Houck B (2024) Toward effective ai support for developers: A survey of desires and concerns. Communications of the ACM 67(11):42--49

\bibitem[{Kitchenham et~al.(2009)Kitchenham, Brereton, Budgen, Turner, Bailey, and Linkman}]{kitchenham2009systematic}
Kitchenham B, Brereton OP, Budgen D, Turner M, Bailey J, Linkman S (2009) Systematic literature reviews in software engineering--a systematic literature review. Information and software technology 51(1):7--15

\bibitem[{Kitchenham et~al.(2010)Kitchenham, Pretorius, Budgen, Brereton, Turner, Niazi, and Linkman}]{kitchenham2010systematic}
Kitchenham B, Pretorius R, Budgen D, Brereton OP, Turner M, Niazi M, Linkman S (2010) Systematic literature reviews in software engineering--a tertiary study. Information and software technology 52(8):792--805

\bibitem[{Klemmer et~al.(2024)Klemmer, Horstmann, Patnaik, Ludden, Burton~Jr, Powers, Massacci, Rahman, Votipka, Lipford et~al.}]{klemmer2024using}
Klemmer JH, Horstmann SA, Patnaik N, Ludden C, Burton~Jr C, Powers C, Massacci F, Rahman A, Votipka D, Lipford HR, et~al. (2024) Using ai assistants in software development: A qualitative study on security practices and concerns. In: Proceedings of the 2024 on ACM SIGSAC Conference on Computer and Communications Security, pp 2726--2740

\bibitem[{Koohestani and Izadi(2025)}]{koohestani2025rethinking}
Koohestani R, Izadi M (2025) Rethinking ide customization for enhanced hax: A hyperdimensional perspective. In: 2025 IEEE/ACM Second IDE Workshop (IDE), IEEE, pp 13--15

\bibitem[{Kruse et~al.(2024)Kruse, Puhlf{\""u}r$\beta$ // this should be~curled beta, and Maalej}]{kruse2024can}
Kruse HA, Puhlf{\""u}r$\beta$ // this should be~curled beta T, Maalej W (2024) Can developers prompt? a controlled experiment for code documentation generation. In: 2024 IEEE International Conference on Software Maintenance and Evolution (ICSME), IEEE, pp 574--586

\bibitem[{Kuang et~al.(2024)Kuang, S{\""o}derberg, and H{\""o}st}]{kuang2024developers}
Kuang P, S{\""o}derberg E, H{\""o}st M (2024) Developers' perspective on today's and tomorrow's programming tool assistance: A survey. In: Companion Proceedings of the 8th International Conference on the Art, Science, and Engineering of Programming, pp 108--116

\bibitem[{Lau and Guo(2023)}]{lau2023ban}
Lau S, Guo P (2023) From"" ban it till we understand it"" to"" resistance is futile"": How university programming instructors plan to adapt as more students use ai code generation and explanation tools such as chatgpt and github copilot. In: Proceedings of the 2023 ACM Conference on International Computing Education Research-Volume 1, pp 106--121

\bibitem[{Li et~al.(2024)Li, Arony, Awon, Damian, and Xu}]{li2024ai}
Li ZS, Arony NN, Awon AM, Damian D, Xu B (2024) Ai tool use and adoption in software development by individuals and organizations: a grounded theory study. arXiv preprint arXiv:240617325

\bibitem[{Liang et~al.(2024)Liang, Yang, and Myers}]{liang2024large}
Liang JT, Yang C, Myers BA (2024) A large-scale survey on the usability of ai programming assistants: Successes and challenges. In: Proceedings of the 46th IEEE/ACM international conference on software engineering, pp 1--13

\bibitem[{Liu et~al.(2024)Liu, Zhang, Zhang, Wan, Huang, and Yan}]{liu2024empirical}
Liu C, Zhang X, Zhang H, Wan Z, Huang Z, Yan M (2024) An empirical study of code search in intelligent coding assistant: Perceptions, expectations, and directions. In: Companion Proceedings of the 32nd ACM International Conference on the Foundations of Software Engineering, pp 283--293

\bibitem[{Manfredi et~al.(2023)Manfredi, Erra, and Gilio}]{manfredi2023mixed}
Manfredi G, Erra U, Gilio G (2023) A mixed reality approach for innovative pair programming education with a conversational ai virtual avatar. In: Proceedings of the 27th International Conference on Evaluation and Assessment in Software Engineering, pp 450--454

\bibitem[{Mastropaolo et~al.(2023)Mastropaolo, Pascarella, Guglielmi, Ciniselli, Scalabrino, Oliveto, and Bavota}]{mastropaolo2023robustness}
Mastropaolo A, Pascarella L, Guglielmi E, Ciniselli M, Scalabrino S, Oliveto R, Bavota G (2023) On the robustness of code generation techniques: An empirical study on github copilot. In: 2023 IEEE/ACM 45th International Conference on Software Engineering (ICSE), IEEE, pp 2149--2160

\bibitem[{McNutt et~al.(2023)McNutt, Wang, Deline, and Drucker}]{mcnutt2023design}
McNutt AM, Wang C, Deline RA, Drucker SM (2023) On the design of ai-powered code assistants for notebooks. In: Proceedings of the 2023 CHI conference on human factors in computing systems, pp 1--16

\bibitem[{Moher et~al.(2010)Moher, Liberati, Tetzlaff, Altman, Group et~al.}]{moher2010preferred}
Moher D, Liberati A, Tetzlaff J, Altman DG, Group P, et~al. (2010) Preferred reporting items for systematic reviews and meta-analyses: the prisma statement. International journal of surgery 8(5):336--341

\bibitem[{de~Moor et~al.(2024)de~Moor, van Deursen, and Izadi}]{de2024transformer}
de~Moor A, van Deursen A, Izadi M (2024) A transformer-based approach for smart invocation of automatic code completion. In: Proceedings of the 1st ACM International Conference on AI-Powered Software, pp 28--37

\bibitem[{Mozannar et~al.(2024{\natexlab{a}})Mozannar, Bansal, Fourney, and Horvitz}]{mozannar2024reading}
Mozannar H, Bansal G, Fourney A, Horvitz E (2024{\natexlab{a}}) Reading between the lines: Modeling user behavior and costs in ai-assisted programming. In: Proceedings of the 2024 CHI Conference on Human Factors in Computing Systems, pp 1--16

\bibitem[{Mozannar et~al.(2024{\natexlab{b}})Mozannar, Bansal, Fourney, and Horvitz}]{mozannar2024show}
Mozannar H, Bansal G, Fourney A, Horvitz E (2024{\natexlab{b}}) When to show a suggestion? integrating human feedback in ai-assisted programming. In: Proceedings of the AAAI Conference on Artificial Intelligence, vol~38, pp 10137--10144

\bibitem[{Mozannar et~al.(2024{\natexlab{c}})Mozannar, Chen, Alsobay, Das, Zhao, Wei, Nagireddy, Sattigeri, Talwalkar, and Sontag}]{mozannar2024realhumaneval}
Mozannar H, Chen V, Alsobay M, Das S, Zhao S, Wei D, Nagireddy M, Sattigeri P, Talwalkar A, Sontag D (2024{\natexlab{c}}) The realhumaneval: Evaluating large language models' abilities to support programmers. arXiv preprint arXiv:240402806

\bibitem[{Nam et~al.(2023)Nam, Macvean, Hellendoorn, Vasilescu, and Myers}]{nam2023ide}
Nam D, Macvean A, Hellendoorn VJ, Vasilescu B, Myers BA (2023) In-ide generation-based information support with a large language model. CoRR

\bibitem[{Nguyen and Nadi(2022)}]{nguyen2022empirical}
Nguyen N, Nadi S (2022) An empirical evaluation of github copilot's code suggestions. In: Proceedings of the 19th International Conference on Mining Software Repositories, pp 1--5

\bibitem[{OBrien et~al.(2024)OBrien, Biswas, Imtiaz, Abdalkareem, Shihab, and Rajan}]{obrien2024prompt}
OBrien D, Biswas S, Imtiaz SM, Abdalkareem R, Shihab E, Rajan H (2024) Are prompt engineering and todo comments friends or foes? an evaluation on github copilot. In: Proceedings of the IEEE/ACM 46th International Conference on Software Engineering, pp 1--13

\bibitem[{Omidvar~Tehrani et~al.(2024)Omidvar~Tehrani, M, and Anubhai}]{omidvar2024evaluating}
Omidvar~Tehrani B, M I, Anubhai A (2024) Evaluating human-ai partnership for llm-based code migration. In: Extended abstracts of the CHI conference on human factors in computing systems, pp 1--8

\bibitem[{Pandey et~al.(2024)Pandey, Singh, Wei, and Shankar}]{pandey2024transforming}
Pandey R, Singh P, Wei R, Shankar S (2024) Transforming software development: Evaluating the efficiency and challenges of github copilot in real-world projects. arXiv preprint arXiv:240617910

\bibitem[{Pearce et~al.(2022)Pearce, Ahmad, Tan, Dolan-Gavitt, and Karri}]{pearce2022asleep}
Pearce H, Ahmad B, Tan B, Dolan-Gavitt B, Karri R (2022) Asleep at the keyboard? assessing the security of github copilot’s code contributions. In: 2022 IEEE Symposium on Security and Privacy (SP), IEEE Computer Society, pp 754--768

\bibitem[{Peng et~al.(2023)Peng, Kalliamvakou, Cihon, and Demirer}]{peng2023impact}
Peng S, Kalliamvakou E, Cihon P, Demirer M (2023) The impact of ai on developer productivity: Evidence from github copilot. arXiv preprint arXiv:230206590

\bibitem[{Penney et~al.(2023)Penney, Pimentel, Steinmacher, and Gerosa}]{penney2023anticipating}
Penney J, Pimentel JF, Steinmacher I, Gerosa MA (2023) Anticipating user needs: Insights from design fiction on conversational agents for computational thinking. In: International Workshop on Chatbot Research and Design, Springer, pp 204--219

\bibitem[{Prather et~al.(2023)Prather, Reeves, Denny, Becker, Leinonen, Luxton-Reilly, Powell, Finnie-Ansley, and Santos}]{prather2023s}
Prather J, Reeves BN, Denny P, Becker BA, Leinonen J, Luxton-Reilly A, Powell G, Finnie-Ansley J, Santos EA (2023) “it’s weird that it knows what i want”: Usability and interactions with copilot for novice programmers. ACM transactions on computer-human interaction 31(1):1--31

\bibitem[{Prather et~al.(2024)Prather, Reeves, Leinonen, MacNeil, Randrianasolo, Becker, Kimmel, Wright, and Briggs}]{prather2024widening}
Prather J, Reeves BN, Leinonen J, MacNeil S, Randrianasolo AS, Becker BA, Kimmel B, Wright J, Briggs B (2024) The widening gap: The benefits and harms of generative ai for novice programmers. In: Proceedings of the 2024 ACM Conference on International Computing Education Research-Volume 1, pp 469--486

\bibitem[{Puryear and Sprint(2022)}]{puryear2022github}
Puryear B, Sprint G (2022) Github copilot in the classroom: learning to code with ai assistance. Journal of Computing Sciences in Colleges 38(1):37--47

\bibitem[{Rasnayaka et~al.(2024)Rasnayaka, Wang, Shariffdeen, and Iyer}]{rasnayaka2024empirical}
Rasnayaka S, Wang G, Shariffdeen R, Iyer GN (2024) An empirical study on usage and perceptions of llms in a software engineering project. In: Proceedings of the 1st International Workshop on Large Language Models for Code, pp 111--118

\bibitem[{Robe and Kuttal(2022)}]{robe2022designing}
Robe P, Kuttal SK (2022) Designing pairbuddy—a conversational agent for pair programming. ACM Transactions on Computer-Human Interaction (TOCHI) 29(4):1--44

\bibitem[{Ross et~al.(2023{\natexlab{a}})Ross, Martinez, Houde, Muller, and Weisz}]{ross2023programmer}
Ross SI, Martinez F, Houde S, Muller M, Weisz JD (2023{\natexlab{a}}) The programmer’s assistant: Conversational interaction with a large language model for software development. In: Proceedings of the 28th International Conference on Intelligent User Interfaces, pp 491--514

\bibitem[{Ross et~al.(2023{\natexlab{b}})Ross, Muller, Martinez, Houde, and Weisz}]{ross2023case}
Ross SI, Muller M, Martinez F, Houde S, Weisz JD (2023{\natexlab{b}}) A case study in engineering a conversational programming assistant's persona. arXiv preprint arXiv:230110016

\bibitem[{S{\'a}godi et~al.(2024)S{\'a}godi, Siket, and Ferenc}]{sagodi2024methodology}
S{\'a}godi Z, Siket I, Ferenc R (2024) Methodology for code synthesis evaluation of llms presented by a case study of chatgpt and copilot. Ieee Access 12:72303--72316

\bibitem[{Sahoo et~al.(2024)Sahoo, Pujar, Nalawade, Genhardt, Mandel, and Buratti}]{sahoo2024ansible}
Sahoo P, Pujar S, Nalawade G, Genhardt R, Mandel L, Buratti L (2024) Ansible lightspeed: A code generation service for it automation. In: Proceedings of the 39th IEEE/ACM International Conference on Automated Software Engineering, pp 2148--2158

\bibitem[{Sandoval et~al.(2023)Sandoval, Pearce, Nys, Karri, Garg, and Dolan-Gavitt}]{sandoval2023lost}
Sandoval G, Pearce H, Nys T, Karri R, Garg S, Dolan-Gavitt B (2023) Lost at c: A user study on the security implications of large language model code assistants. In: 32nd USENIX Security Symposium (USENIX Security 23), pp 2205--2222

\bibitem[{Senanayake et~al.(2024)Senanayake, Kalutarage, Petrovski, Piras, and Al-Kadri}]{senanayake2024defendroid}
Senanayake J, Kalutarage H, Petrovski A, Piras L, Al-Kadri MO (2024) Defendroid: Real-time android code vulnerability detection via blockchain federated neural network with xai. Journal of Information Security and Applications 82:103741

\bibitem[{Sergeyuk et~al.(2024)Sergeyuk, Titov, and Izadi}]{sergeyuk2024ide}
Sergeyuk A, Titov S, Izadi M (2024) In-ide human-ai experience in the era of large language models; a literature review. In: Proceedings of the 1st ACM/IEEE Workshop on Integrated Development Environments, pp 95--100

\bibitem[{Sergeyuk et~al.(2025)Sergeyuk, Zakharov, Koshchenko, and Izadi}]{hax_dataset_2025}
Sergeyuk A, Zakharov I, Koshchenko E, Izadi M (2025) Dataset for the systematic literature review on in-ide human--ai experience. \doi{10.5281/zenodo.16877797}, \urlprefix\url{https://doi.org/10.5281/zenodo.16877797}, version v2

\bibitem[{Shlomov et~al.(2024)Shlomov, Yaeli, Marreed, Schwartz, Eder, Akrabi, and Zeltyn}]{shlomov2024ida}
Shlomov S, Yaeli A, Marreed S, Schwartz S, Eder N, Akrabi O, Zeltyn S (2024) Ida: Breaking barriers in no-code ui automation through large language models and human-centric design. arXiv preprint arXiv:240715673

\bibitem[{Spiess et~al.(2025)Spiess, Gros, Pai, Pradel, Rabin, Alipour, Jha, Devanbu, and Ahmed}]{spiess2025calibration}
Spiess C, Gros D, Pai KS, Pradel M, Rabin MRI, Alipour A, Jha S, Devanbu P, Ahmed T (2025) Calibration and correctness of language models for code. In: 2025 IEEE/ACM 47th International Conference on Software Engineering (ICSE), IEEE, pp 540--552

\bibitem[{{Stack Overflow}(2024)}]{StackOverflow2024AI}
{Stack Overflow} (2024) {2024 Stack Overflow Developer Survey: AI}. \urlprefix\url{https://survey.stackoverflow.co/2024/ai}, accessed: 2025-08-10

\bibitem[{Sun et~al.(2022)Sun, Liao, Muller, Agarwal, Houde, Talamadupula, and Weisz}]{sun2022investigating}
Sun J, Liao QV, Muller M, Agarwal M, Houde S, Talamadupula K, Weisz JD (2022) Investigating explainability of generative ai for code through scenario-based design. In: Proceedings of the 27th international conference on intelligent user interfaces, pp 212--228

\bibitem[{Süße et~al.(2023)Süße, Kobert, Grapenthin, and Voigt}]{susse2023ai}
Süße T, Kobert M, Grapenthin S, Voigt BF (2023) Ai-powered chatbots and the transformation of work: Findings from a case study in software development and software engineering. In: Working Conference on Virtual Enterprises, Springer, pp 689--705

\bibitem[{Tan et~al.(2023)Tan, Guo, Wong, and Hang}]{tan2023copilot}
Tan CW, Guo S, Wong MF, Hang CN (2023) Copilot for xcode: exploring ai-assisted programming by prompting cloud-based large language models. arXiv preprint arXiv:230714349

\bibitem[{Tanay et~al.(2024)Tanay, Arinze, Joshi, Davis, and Davis}]{tanay2024exploratory}
Tanay BA, Arinze L, Joshi SS, Davis KA, Davis JC (2024) An exploratory study on upper-level computing students' use of large language models as tools in a semester-long project. arXiv preprint arXiv:240318679

\bibitem[{Tang et~al.(2024)Tang, Chen, Ning, Bansal, Huang, McMillan, and Li}]{tang2024study}
Tang N, Chen M, Ning Z, Bansal A, Huang Y, McMillan C, Li TJJ (2024) A study on developer behaviors for validating and repairing llm-generated code using eye tracking and ide actions. arXiv preprint arXiv:240516081

\bibitem[{Tian et~al.(2023)Tian, Zhang, Ning, Li, Kummerfeld, and Zhang}]{tian2023interactive}
Tian Y, Zhang Z, Ning Z, Li TJJ, Kummerfeld JK, Zhang T (2023) Interactive text-to-sql generation via editable step-by-step explanations. In: EMNLP

\bibitem[{Tian et~al.(2024)Tian, Kummerfeld, Li, and Zhang}]{tian2024sqlucid}
Tian Y, Kummerfeld JK, Li TJJ, Zhang T (2024) Sqlucid: Grounding natural language database queries with interactive explanations. In: Proceedings of the 37th Annual ACM Symposium on User Interface Software and Technology, pp 1--20

\bibitem[{Tong and Zhang(2024)}]{tong2024codejudge}
Tong W, Zhang T (2024) Codejudge: Evaluating code generation with large language models. In: Proceedings of the 2024 Conference on Empirical Methods in Natural Language Processing, pp 20032--20051

\bibitem[{Vaithilingam et~al.(2022)Vaithilingam, Zhang, and Glassman}]{vaithilingam2022expectation}
Vaithilingam P, Zhang T, Glassman EL (2022) Expectation vs. experience: Evaluating the usability of code generation tools powered by large language models. In: Chi conference on human factors in computing systems extended abstracts, pp 1--7

\bibitem[{Vaithilingam et~al.(2023)Vaithilingam, Glassman, Groenwegen, Gulwani, Henley, Malpani, Pugh, Radhakrishna, Soares, Wang et~al.}]{vaithilingam2023towards}
Vaithilingam P, Glassman EL, Groenwegen P, Gulwani S, Henley AZ, Malpani R, Pugh D, Radhakrishna A, Soares G, Wang J, et~al. (2023) Towards more effective ai-assisted programming: A systematic design exploration to improve visual studio intellicode’s user experience. In: 2023 IEEE/ACM 45th International Conference on Software Engineering: Software Engineering in Practice (ICSE-SEIP), IEEE, pp 185--195

\bibitem[{Vasconcelos et~al.(2025)Vasconcelos, Bansal, Fourney, Liao, and Wortman~Vaughan}]{vasconcelos2025generation}
Vasconcelos H, Bansal G, Fourney A, Liao QV, Wortman~Vaughan J (2025) Generation probabilities are not enough: Uncertainty highlighting in ai code completions. ACM Transactions on Computer-Human Interaction 32(1):1--30

\bibitem[{Vasiliniuc and Groza(2023)}]{vasiliniuc2023case}
Vasiliniuc MS, Groza A (2023) Case study: using ai-assisted code generation in mobile teams. In: 2023 IEEE 19th International Conference on Intelligent Computer Communication and Processing (ICCP), IEEE, pp 339--346

\bibitem[{Wang et~al.(2022)Wang, Wang, Drozdal, Muller, Park, Weisz, Liu, Wu, and Dugan}]{wang2022documentation}
Wang AY, Wang D, Drozdal J, Muller M, Park S, Weisz JD, Liu X, Wu L, Dugan C (2022) Documentation matters: Human-centered ai system to assist data science code documentation in computational notebooks. ACM Transactions on Computer-Human Interaction 29(2):1--33

\bibitem[{Wang et~al.(2023{\natexlab{a}})Wang, Hu, Gao, Jin, Xie, Huang, Lei, and Deng}]{wang2023practitioners}
Wang C, Hu J, Gao C, Jin Y, Xie T, Huang H, Lei Z, Deng Y (2023{\natexlab{a}}) Practitioners' expectations on code completion. arXiv preprint arXiv:230103846

\bibitem[{Wang et~al.(2023{\natexlab{b}})Wang, Liu, Liu, Neshati, Ma, Zhu, and Zhao}]{wang2023slide4n}
Wang F, Liu X, Liu O, Neshati A, Ma T, Zhu M, Zhao J (2023{\natexlab{b}}) Slide4n: Creating presentation slides from computational notebooks with human-ai collaboration. In: Proceedings of the 2023 CHI Conference on Human Factors in Computing Systems, pp 1--18

\bibitem[{Wang et~al.(2024)Wang, Cheng, Ford, and Zimmermann}]{wang2024investigating}
Wang R, Cheng R, Ford D, Zimmermann T (2024) Investigating and designing for trust in ai-powered code generation tools. In: Proceedings of the 2024 ACM Conference on Fairness, Accountability, and Transparency, pp 1475--1493

\bibitem[{Wang et~al.(2025)Wang, Zhou, Song, Huang, Chen, Ma, and Zhang}]{wang2025towards}
Wang Z, Zhou Z, Song D, Huang Y, Chen S, Ma L, Zhang T (2025) Towards understanding the characteristics of code generation errors made by large language models. In: 2025 IEEE/ACM 47th International Conference on Software Engineering (ICSE), IEEE Computer Society, pp 717--717

\bibitem[{Weber et~al.(2024)Weber, Brandmaier, Schmidt, and Mayer}]{weber2024significant}
Weber T, Brandmaier M, Schmidt A, Mayer S (2024) Significant productivity gains through programming with large language models. Proceedings of the ACM on Human-Computer Interaction 8(EICS):1--29

\bibitem[{Weisz et~al.(2022)Weisz, Muller, Ross, Martinez, Houde, Agarwal, Talamadupula, and Richards}]{weisz2022better}
Weisz JD, Muller M, Ross SI, Martinez F, Houde S, Agarwal M, Talamadupula K, Richards JT (2022) Better together? an evaluation of ai-supported code translation. In: Proceedings of the 27th International Conference on Intelligent User Interfaces, pp 369--391

\bibitem[{Wermelinger(2023)}]{wermelinger2023using}
Wermelinger M (2023) Using github copilot to solve simple programming problems. In: Proceedings of the 54th ACM Technical Symposium on Computer Science Education V. 1, pp 172--178

\bibitem[{Wohlin(2014)}]{wohlin2014guidelines}
Wohlin C (2014) Guidelines for snowballing in systematic literature studies and a replication in software engineering. In: Proceedings of the 18th international conference on evaluation and assessment in software engineering, pp 1--10

\bibitem[{Yan et~al.(2024)Yan, Hwang, Wu, and Head}]{yan2024ivie}
Yan L, Hwang A, Wu Z, Head A (2024) Ivie: Lightweight anchored explanations of just-generated code. In: Proceedings of the 2024 CHI Conference on Human Factors in Computing Systems, pp 1--15

\bibitem[{Yen et~al.(2023)Yen, Zhu, Suh, Xia, and Zhao}]{yen2023coladder}
Yen R, Zhu J, Suh S, Xia H, Zhao J (2023) Coladder: Supporting programmers with hierarchical code generation in multi-level abstraction. arXiv preprint arXiv:231008699

\bibitem[{Yetistiren et~al.(2022)Yetistiren, Ozsoy, and Tuzun}]{yetistiren2022assessing}
Yetistiren B, Ozsoy I, Tuzun E (2022) Assessing the quality of github copilot’s code generation. In: Proceedings of the 18th international conference on predictive models and data analytics in software engineering, pp 62--71

\bibitem[{Zhang et~al.(2023)Zhang, Liang, Zhou, Ahmad, and Waseem}]{zhang2023practices}
Zhang B, Liang P, Zhou X, Ahmad A, Waseem M (2023) Practices and challenges of using github copilot: An empirical study. arXiv preprint arXiv:230308733

\bibitem[{Zhou and Li(2023)}]{zhou2023case}
Zhou H, Li J (2023) A case study on scaffolding exploratory data analysis for ai pair programmers. In: Extended Abstracts of the 2023 CHI Conference on Human Factors in Computing Systems, pp 1--7

\bibitem[{Zhou et~al.(2025{\natexlab{a}})Zhou, Cao, Sun, and Lo}]{zhou2025large}
Zhou X, Cao S, Sun X, Lo D (2025{\natexlab{a}}) Large language model for vulnerability detection and repair: Literature review and the road ahead. ACM Transactions on Software Engineering and Methodology 34(5):1--31

\bibitem[{Zhou et~al.(2025{\natexlab{b}})Zhou, Liang, Zhang, Li, Ahmad, Shahin, and Waseem}]{zhou2025exploring}
Zhou X, Liang P, Zhang B, Li Z, Ahmad A, Shahin M, Waseem M (2025{\natexlab{b}}) Exploring the problems, their causes and solutions of ai pair programming: A study on github and stack overflow. Journal of Systems and Software 219:112204

\bibitem[{Zhou et~al.(2015)Zhou, Zhang, Huang, Yang, Babar, and Tang}]{zhou2015quality}
Zhou Y, Zhang H, Huang X, Yang S, Babar MA, Tang H (2015) Quality assessment of systematic reviews in software engineering: A tertiary study. In: Proceedings of the 19th international conference on evaluation and assessment in software engineering, pp 1--14

\bibitem[{Zhu et~al.(2024)Zhu, Wang, Ma, Wang, Chen, Khurana, and Ma}]{zhu2024towards}
Zhu Q, Wang D, Ma S, Wang AY, Chen Z, Khurana U, Ma X (2024) Towards feature engineering with human and ai’s knowledge: Understanding data science practitioners’ perceptions in human\&ai-assisted feature engineering design. In: Proceedings of the 2024 ACM Designing Interactive Systems Conference, pp 1789--1804

\bibitem[{Ziegler et~al.(2022)Ziegler, Kalliamvakou, Li, Rice, Rifkin, Simister, Sittampalam, and Aftandilian}]{ziegler2022productivity}
Ziegler A, Kalliamvakou E, Li XA, Rice A, Rifkin D, Simister S, Sittampalam G, Aftandilian E (2022) Productivity assessment of neural code completion. In: Proceedings of the 6th ACM SIGPLAN International Symposium on Machine Programming, pp 21--29

\end{thebibliography}

\end{document}